\documentclass[aps,superscriptaddress,notitlepage,onecolumn,bibnotes,
]{revtex4-1}
\usepackage[utf8]{inputenc}
\usepackage{graphicx}
\usepackage{float}
\usepackage{gensymb}
\usepackage{upgreek}
\usepackage{color,amsmath}
\usepackage[colorlinks,linkcolor=blue,citecolor=blue,urlcolor=blue]{hyperref}
\usepackage[normalem]{ulem}
\usepackage{siunitx}
\usepackage{enumitem}
\usepackage{ulem}
\usepackage{amsmath,amsfonts}
 
\usepackage{ragged2e}
\usepackage{textcomp}
\usepackage{color,soul}
\usepackage{textgreek}
\usepackage{hyphenat}
\usepackage{adjustbox}
\usepackage{setspace}
\usepackage{xcolor}
\usepackage{soul}
\usepackage{fancyhdr}
\usepackage{graphicx}
\usepackage{lineno}
\colorlet{lightyellow}{yellow!40}

\sethlcolor{lightyellow}

\usepackage{natbib}
\setcitestyle{numbers}  
\usepackage{natbib}

\renewcommand{\cite}[1]{\textsuperscript{\citenum{#1}}}

\renewcommand{\doi}[1]{}
\renewcommand{\url}[1]{}

\makeatletter
\renewcommand\section{\@startsection {section}{1}{\z@}%
   {-2.0ex \@plus -1ex \@minus -.2ex}%
   {1.0ex \@plus.2ex}%
   {\normalfont\bfseries}}
\renewcommand\subsection{\@startsection {subsection}{2}{\z@}%
   {-1.5ex \@plus -1ex \@minus -.2ex}%
   {0.5ex \@plus .2ex}%
   {\normalfont\bfseries}}
\makeatother

\DeclareUnicodeCharacter{2212}{-}

\makeatother

\begin{document}
\setlength{\parskip}{0.55em}  

\title{\large Control of polarization and polar chiral textures in BiFeO\textsubscript{3}\\ by epitaxial strain and interfacial chemistry}



\author{Elzbieta Gradauskaite}
\email{elzbieta.gradauskaite@cnrs-thales.fr}
\affiliation{Laboratoire Albert Fert, CNRS, Thales, Universit\'e Paris--Saclay, Palaiseau, France}
\affiliation{Department of Materials, ETH Zurich, Zurich, Switzerland}

\author{Natascha Gray}
\affiliation{Department of Materials, ETH Zurich, Zurich, Switzerland}

\author{Quintin N. Meier}
\affiliation{Universit\'e Grenoble Alpes, CNRS, Institut N\'eel, Grenoble, France}

\author{Marta D. Rossell}
\affiliation{Electron Microscopy Center, Empa, D\"ubendorf, Switzerland}

\author{Morgan Trassin}
\email{morgan.trassin@mat.ethz.ch}
\affiliation{Department of Materials, ETH Zurich, Zurich, Switzerland}

\begin{abstract}
The balance between interfacial chemistry, electrostatics, and epitaxial strain plays a crucial role in stabilizing polarization in ferroelectric thin films. Here, we bring these contributions into competition in BiFeO\textsubscript{3} (BFO) thin films grown on the charged-surface-terminated La\textsubscript{0.7}Sr\textsubscript{0.3}MnO\textsubscript{3} (LSMO)-buffered NdGaO\textsubscript{3} (001) substrates. The large anisotropic compressive strain from the substrate promotes the formation of ferroelectric domains despite the expected stabilization of a uniform out-of-plane polarization by the (La,Sr)O\textsuperscript{0.7+} termination of the metallic buffer. Piezoresponse force microscopy and scanning transmission electron microscopy reveal that the resulting nanoscale domain architecture is stabilized by the deterministic formation of a fluorite-like Bi\textsubscript{2}O\textsubscript{2} surface layer on regions polarized oppositely to the LSMO-imposed polarization orientation. Leveraging this polarization compensation mechanism, we stabilize a uniform out-of-plane polarization in our highly strained BFO films by inserting a Bi\textsubscript{2}O\textsubscript{2}-terminated Aurivillius film as a buffer layer. Additionally, we reveal signatures of homochiral polarization textures in our BFO films on the level of domain configurations using local polarization switching experiments. Our work thus brings new strategies for controlling polarization direction and chiral textures in oxide ferroelectrics, opening pathways for functional domain-wall and domain-based electronics.
\end{abstract}

\maketitle

\section{Introduction}

Ferroelectric thin films are promising candidates for next-generation electronic devices due to their non-volatile polarization, which can be switched solely using electric fields\cite{Scott2000book}, making them attractive as energy-efficient alternatives to conventional complementary metal-oxide-semiconductor (CMOS)-based technologies\cite{muller2023}. However, the integration of ferroelectrics into functional devices requires precise control over their polarization state, particularly at the nanoscale, where depolarizing fields\cite{Lichtensteiger2014a, Strkalj2019}, surface chemistry\cite{Strkalj2020, efe2024}, epitaxial strain\cite{Zeches2009, Damodaran2014,nordlander2024}, and defects\cite{Li2018a, Weymann2020, gradauskaite2022, sarott2023} strongly impact the final polarization configuration. These factors collectively dictate the stable polarization components, domain structures, and the types and textures of domain walls, which all influence the performance of ferroelectric devices. 

In thin-film perovskite-based ferroelectrics, the out-of-plane polarization component is typically controlled through engineering of top and bottom interfaces, usually via the choice of electrodes\cite{Yu2012a} or insulating buffer layers\cite{Lichtensteiger2014a}. At these interfaces, the electrostatic boundary conditions determine whether a uniform or multidomain state forms and, in extreme cases, can even lead to the complete suppression of polarization\cite{Junquera2003, Kim2005a}. Specifically, electrode surface terminations with charged atomic planes have been used to impose a preferred out-of-plane polarization direction\cite{Yu2012a}. Meanwhile, the formation of in-plane-polarized domains is primarily dictated by epitaxial strain, through both lattice\cite{Schlom2007} and symmetry mismatch\cite{Chen2014} imposed by the substrate. For instance, tensile epitaxial strain can increase\cite{catalan2006a} or even induce in-plane polar distortions\cite{Becher2015a}, while anisotropic in-plane strain can give rise to domain architectures with preferential in-plane polarization components\cite{Chu2009}. To date, the combination of tensile epitaxial strain and interface depolarizing fields has been mainly employed to enforce nanoscale domain architectures in oxide superlattices. In particular, the finely tuned balance between elastic, electrostatic, and gradient energy terms in the ferroelectric\textbar dielectric superlattices can even stabilize unconventional polar textures, such as flux-closure domains\cite{Tang2015}, polar vortices\cite{Yadav2016,Hong2017}, or skyrmions\cite{Das2019a}. However, the interplay between epitaxial strain, which promotes domain formation, and interfacial electrostatic effects that favor a single-domain state has remained underexplored. In such cases, the accommodation of epitaxial strain and the preferential screening of bound polarization charges cannot be achieved simultaneously. The resulting frustration could then be harnessed for novel design\cite{yin2024} of functional ferroelectric, antiferroelectric, and multiferroic domains and domain walls.

Here,  we investigate the competition between epitaxial strain, which favors multidomain formation, and interfacial electrostatics supporting a single out-of-plane polarization component, in multiferroic BiFeO\textsubscript{3} (BFO) thin films. We use NdGaO\textsubscript{3} (NGO) (001)-oriented substrates to impose anisotropic compressive strain and promote in-plane and out-of-plane polarized ferroelastic domain formation in our films. To create a conflicting electrostatic condition, we insert a (La,Sr)O\textsuperscript{0.7+}-terminated La\textsubscript{0.7}Sr\textsubscript{0.3}MnO\textsubscript{3} (LSMO) buffer, which favors the growth of films with uniform downward-oriented polarization. Using a combination of piezoresponse force microscopy (PFM) and scanning transmission electron microscopy (STEM), we study the consequences of the frustration arising from the competing elastic and electrostatic boundary conditions and reveal the onset of a multidomain state in both in-plane and out-of-plane polarization components in our BFO films. A deterministic formation of Bi\(_2\)O\(_2\) surface layer occurs in regions where the polarization opposes the electrostatic constraints imposed by the charged termination of the metallic buffer, in this way providing a bound-charge screening from the top surface.  Building on this observation, we use the naturally forming Bi\textsubscript{2}O\textsubscript{2} surface layer of the Aurivillius buffer layer to stabilize domain variants exhibiting a uniform out-of-plane polarization in highly strained BFO on NGO (001). Finally, in this system, we reveal signatures of the recently reported polar chiral textures in BFO at the macroscopic level with the generation of deterministic domain ordering following local polarization switching experiments. By elucidating the interplay between elastic and electrostatic boundary conditions, we establish a pathway for engineering ferroelectric domain and domain-wall configurations with a defined chirality in oxide heterostructures.

\section{Results and Discussion} 

\subsection{Competing polarization constraints in ultrathin BFO films on LSMO-buffered NGO (001)}

\begin{figure}[htb!]
  \centering 
  \includegraphics[width=0.85\textwidth, clip, trim=4 4 4 4]{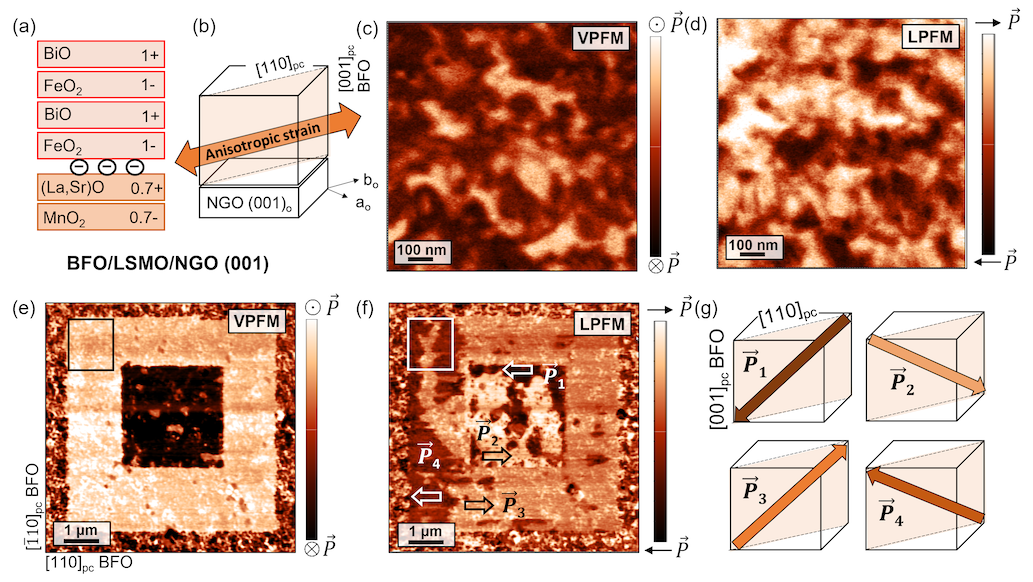}
  \caption{\textbf{Competing strain and electrostatic constraints in BFO thin films on LSMO-buffered NGO (001) that lead to a nanoscale multidomain state.}  
  (a) Schematic of the atomic \textit{A}O and \textit{B}O\textsubscript{2} planes in a (La,Sr)O\textsuperscript{0.7+}-terminated LSMO and BFO, illustrating the net negative charge at the interface, which favors a downward polarization state.  
  (b) The NGO (001) substrate imposes a highly anisotropic in-plane compressive strain on BFO, with strain components of -3.2\% and -1.9\% along its orthorhombic in-plane $a$\textsubscript{o} and $b$\textsubscript{o}-axes, respectively. This favors a uniaxial in-plane polarization anisotropy in BFO along [110]\textsubscript{pc} parallel to the $b$\textsubscript{o}-axis of NGO.  (c, d) VPFM (c) and LPFM (d) images of pristine BFO reveal a nanoscale multidomain state despite the electrostatic preference for uniform out-of-plane polarization. (e, f) PFM images of artificially poled regions: (e) VPFM confirms the formation of larger upward- and downward-polarized domains, while (f) LPFM reveals the corresponding two in-plane polarization components associated with each up and down domain.  
  (g) Schematic of the four polarization variants present in the pristine film ($\vec{P}$\textsubscript{1}, $\vec{P}$\textsubscript{2}, $\vec{P}$\textsubscript{3}, and $\vec{P}$\textsubscript{4}) constrained to the $(\overline{1}10)_{\mathrm{pc}}$ plane of BFO.}  
  \label{fig:1}
\end{figure}

In this study, we induce competition between electrostatic interfacial effects and epitaxial strain in determining the polarization state of BFO thin films by growing them on an LSMO buffer layer deposited on NGO (001) substrates. In epitaxial BFO, a uniform out-of-plane polarization can be enforced by controlling the surface termination\cite{rijnders04} of the underlying buffer layer\cite{Yu2012a,deluca2017a, gradauskaite2024}. Here, we use a (La,Sr)O\textsuperscript{0.7+}-terminated LSMO layer to promote downward-oriented polarization in pseudocubic (001)-oriented BFO films (Fig.~\ref{fig:1}a). At the interface, the (La,Sr)O\textsuperscript{0.7+}/FeO\textsuperscript{1–} plane sequence generates a net negative charge, stabilizing a downward polarization state in BFO. The LSMO also serves as a bottom electrode for PFM measurements. Meanwhile, the compressive epitaxial strain imposed by NGO (001) is highly anisotropic, with -3.2\% and -1.9\% strain along the $a$- and $b$-axes of the orthorhombic NGO substrate ($a_o$, $b_o$), respectively. This strain state selectively stabilizes in-plane polarization components along the less compressed [110]\textsubscript{pc} axis of BFO aligned parallel to $b_o$, as schematized in Figure \ref{fig:1}b. This highly constrained elastic environment with a uniaxial in-plane polarization favors multidomain nucleation.

Lateral PFM (LPFM) and vertical PFM (VPFM) measurements (Fig.~\ref{fig:1}c, d) on our 25-nm-thick, coherently strained BFO films on LSMO/NGO (001) reveal nanoscale domains, giving the first indication that the strain-induced uniaxial in-plane polar anisotropy is competing with the out-of-plane polarization imposed by the charged LSMO surface. The domains observed in the in-plane and out-of-plane PFM channels only partially overlap, suggesting a complex nanoscale domain configuration with independently varying in-plane and out-of-plane polarization components. To shed light on the existing ferroelectric domain variants present in our films, we performed local poling with the AFM tip to create larger upward- and downward-polarized regions (Fig.~\ref{fig:1}e). The poled areas exhibit solely two in-plane polarization components (Fig.~\ref{fig:1}f), oriented along the longer $b_o$-axis of the orthorhombic NGO substrate (corresponding to [110]\textsubscript{pc} of BFO), consistent with the expected anisotropic strain\cite{gradauskaite2023,abdelsamie2024} (see Fig.\ S1, Supporting Information, for vector PFM measurements). As both upward- and downward-polarized domains appear in the pristine film (Fig.\ \ref{fig:1}c), we conclude that four different domain variants are present in the as-grown BFO film, labeled as $\vec{P}$\textsubscript{1}, $\vec{P}$\textsubscript{2}, $\vec{P}$\textsubscript{3}, and $\vec{P}$\textsubscript{4} in Figure~\ref{fig:1}g.

\subsection{Nanodomain formation compensated by Bi\textsubscript{2}O\textsubscript{2} surface reconstructions in BFO}


To rationalize domain nucleation in our BFO films on surface-terminated LSMO buffer layers, we first consider the elastic boundary conditions. The NGO (001) substrate imposes a large, anisotropic in-plane misfit strain. The film relieves this strain by forming two monoclinic BFO variants whose polarization vectors lie in the (100)\textsubscript{o} plane of NGO\cite{abdelsamie2024} ($(\overline{1}10)_{\mathrm{pc}}$ BFO plane, Fig.~\ref{fig:1}b). This leads to an alternation of the in-plane polarization along the NGO $b_o$-axis ([110]\textsubscript{pc} axis of BFO). At the same time, the positively charged surface of the LSMO buffer favors a uniform downward out-of-plane polarization\cite{Yu2012a}, setting up an electrostatic preference in opposition to the strain-driven domain formation. Satisfying both requirements simultaneously would create electrostatically costly charged head-to-head or tail-to-tail domain walls\cite{campanini2020a,Nataf2020}. Local reversals of the out-of-plane polarization in the films, hence, lower the system's energy as charged domain walls (CDWs) are avoided in the final nanoscale ferroelectric domain architecture.

\begin{figure}[htb!]
  \centering 
  \begin{adjustbox}{width=0.75\textwidth, center}
    \includegraphics[width=0.75\textwidth, clip, trim=8 8 8 8]{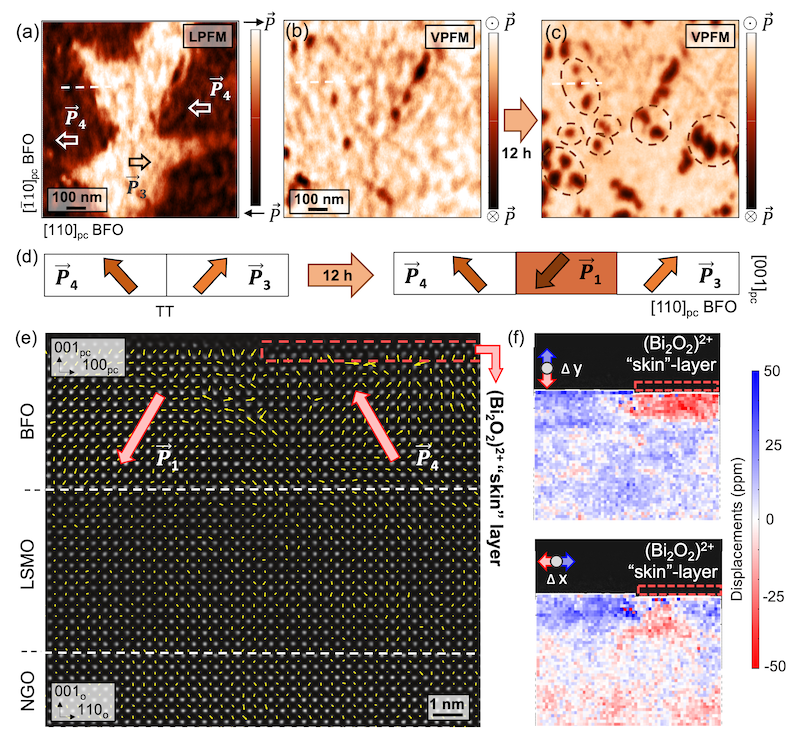}
  \end{adjustbox}
  \caption{\textbf{Instability of charged domain walls (CDWs) and Bi\textsubscript{2}O\textsubscript{2}-based compensation of upward-polarized domains in BFO.}
(a) LPFM and (b) VPFM phase images of the boxed area in Fig.~\ref{fig:1}e,f after upward poling, showing head-to-head and tail-to-tail CDWs between the $\vec{P}_4$ and $\vec{P}_3$ variants.  
(c) Twelve hours later, VPFM reveals a spontaneous switch-back to the downward state along the CDW contours, confirming their electrostatic instability. (d) Schematic cross-section through the film (along the dashed line in a–c, i.e., $(\overline{1}10)_{\mathrm{pc}}$ BFO plane), illustrating projected polarization components of the tail-to-tail domain wall before and after the switch-back. (e) HAADF-STEM image of the BFO/LSMO/NGO heterostructure with overlaid ferroelectric dipole map, revealing nanoscale domains with reversed out-of-plane polarization components ($\vec{P}_1$ and $\vec{P}_4$) in the pristine film as well as a fluorite-like Bi\textsubscript{2}O\textsubscript{2} surface layer forming over a fraction of the film (region outlined in red). (f) Mapping of the out-of-plane (top) and in-plane (bottom) atomic \textit{B}-site displacements extracted from (e) identifies two domains, with the Bi\textsubscript{2}O\textsubscript{2} layer forming exclusively over the upward domain to provide the charge needed to stabilize this otherwise unfavorable polarization direction. Note that the ferroelectric dipoles in (e) are shown opposite to the displacement of the $B$ cations in (f).}\label{fig:2}
\end{figure}

The system’s resistance to a uniform out-of-plane polarization, linked to the undesirable formation of CDWs, is evident in local poling experiments. In the regions poled upward, marked in Figure \ref{fig:1}e,f, and zoomed-in LPFM and VPFM images in Figure \ref{fig:2}a,b, head-to-head and tail-to-tail walls between the $\vec{P}_4$ and $\vec{P}_3$ variants can be initially identified right after the local bias application. A follow-up VPFM scan acquired 12 hours later (Fig.\ \ref{fig:2}c) shows that the walls have spontaneously reverted to the downward-polarized state. This rapid switch-back confirms that CDWs are energetically unstable, prompting the film to lower its energy by locally inverting the out-of-plane polarization component. Figure \ref{fig:2}d shows a cross-sectional representation of the polarization components associated with domain configurations before and after the spontaneous switch-back at the charged tail-to-tail domain wall.

To pinpoint the mechanism that stabilizes the multidomain state and, in particular, the \textit{a priori} unfavorable upward-polarized domains, which override the electrostatic preference of the LSMO electrode, we performed high-angle annular dark-field scanning transmission electron microscopy (HAADF-STEM) analysis of the BFO/LSMO/NGO heterostructure (Fig.~\ref{fig:2}e). Two distinct polarization states are captured within a single field of view, corresponding to $\vec{P}_4$ (up) and $\vec{P}_1$ (down) domains. At the surface of the film, we observe an atomic reconstruction (red box) that coincides precisely and exclusively with the upward-polarized regions, as revealed by atomically resolved mapping of out-of-plane and in-plane $B$-site cation displacements in Figure \ref{fig:2}f. We identify this ``skin layer'' as a fluorite-like Bi\textsubscript{2}O\textsubscript{2} reconstruction that acts as a surface compensation mechanism, mitigating the lack of screening for upward-polarized domains from the bottom interface. These Bi\textsubscript{2}O\textsubscript{2} layers comprise a negatively charged double-oxygen plane and can therefore preferentially stabilize upward-oriented domains. Such layers have been previously identified as polarizing layers in BFO thin films\cite{jin2017b,Spaldin2021,Xie2017,efe2025}. Here, we directly correlate their formation with the underlying polarization direction of the film. We note that the presence of the Bi\textsubscript{2}O\textsubscript{2} ``skin layers'' also likely accounts for the patchy contrast modulations observed in the piezoresponse signal of our BFO film.

\subsection{Deterministic insertion of Bi\textsubscript{2}O\textsubscript{2} layers to enforce a uniform out-of-plane polarization in BFO}

\begin{figure}[htb!]
  \centering 
  \begin{adjustbox}{width=0.8\textwidth, center}
    \includegraphics[width=0.8\textwidth, clip, trim=7 7 7 7]{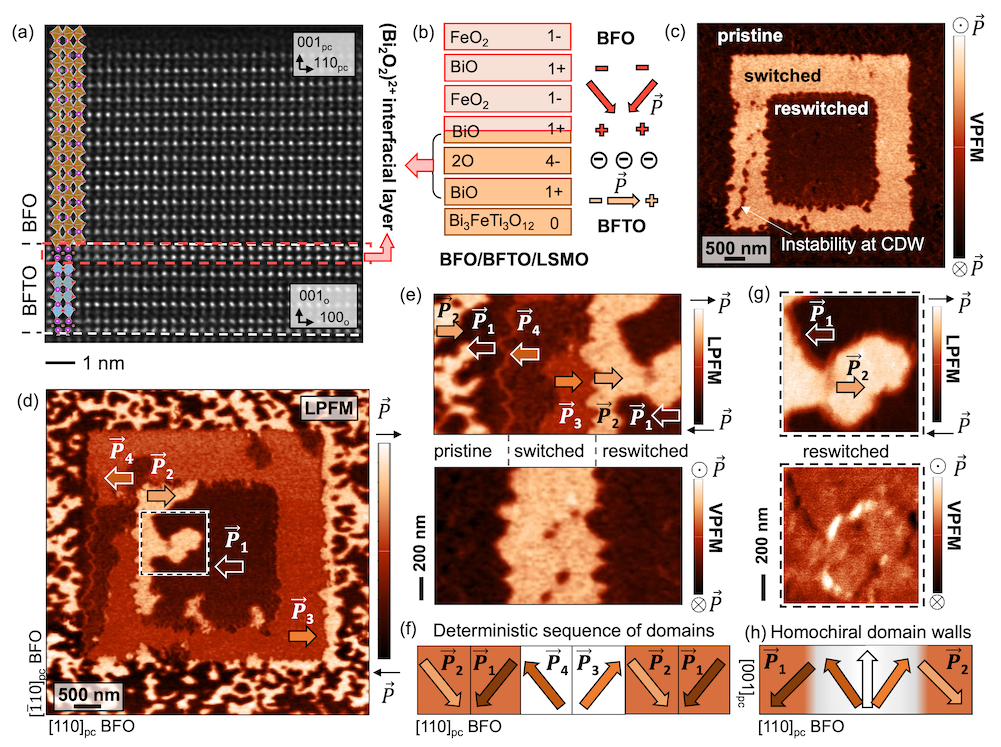}
  \end{adjustbox}
  \caption{\textbf{Electrostatic stabilization of uniform out-of-plane polarization and homochiral polarization configurations in BFO with an interfacial Bi\textsubscript{2}O\textsubscript{2} layer.}  
   (a) HAADF-STEM image of the BFO/BFTO/LSMO heterostructure, revealing an atomically sharp interface between the Aurivillius $n$=4 compound BFTO and BFO. The insertion of one unit cell of BFTO introduces a Bi\textsubscript{2}O\textsubscript{2} interfacial layer (highlighted with a dashed red box).  (b) Schematic of the charged atomic planes that make up the BFO/BFTO interface, highlighting the unscreened negative charges associated with the Bi\textsubscript{2}O\textsubscript{2} interfacial layer that favor a downward-pointing polarization. (c) VPFM confirms that as-grown BFO on BFTO/NGO is uniformly downward-polarized, while poling allows reversible switching between up- and downward-polarized states. (d) LPFM reveals that in the pristine film, only two polarization components $\vec{P}_1$ and $\vec{P}_2$ are observed, with $\vec{P}_3$ and $\vec{P}_4$ appearing when polarization is poled upward. (e) LPFM (top) and VPFM (bottom) images of pristine, switched, and reswitched areas of the film that adopt deterministic domain configurations, as schematically illustrated in (f).
  (g) LPFM (top) and VPFM (bottom) images of the reswitched domains with homochiral polarization rotation apparent at the domain walls. (h) Schematic of polarization rotation at a tail-to-tail domain wall consistent with (g).}
  \label{fig:4}
\end{figure}

The STEM image in Figure \ref{fig:2}e, together with previous reports\cite{jin2017b,Spaldin2021}, suggests that an insulating Bi\textsubscript{2}O\textsubscript{2} “skin layer” may be more effective in enforcing a uniform out-of-plane polarization in BFO under anisotropic epitaxial strain than conventional metallic perovskite buffers with charged surface terminations. Motivated by this, we propose a strategy to stabilize a uniform polarization state in BFO by introducing a well-controlled and uniform Bi\textsubscript{2}O\textsubscript{2} interlayer beneath the film. Such fluorite-like Bi\textsubscript{2}O\textsubscript{2} layers form spontaneously as part of the Aurivillius crystal structure\cite{aurivilliusmain,Moure2018a,gradauskaite2020,Gradauskaite2021a}. In our previous work, we demonstrated the structural compatibility between Bi\textsubscript{3}FeTi\textsubscript{5}O\textsubscript{15} (BFTO) Aurivillius compounds and BFO\cite{efe2025} and showed that in-plane-polarized Aurivillius buffers can suppress depolarizing field effects in ultrathin ferroelectric films\cite{gradauskaite2023}. Here, we focus on how Bi\textsubscript{2}O\textsubscript{2} layers could influence the domain architecture in BFO films grown on LSMO-buffered NGO (001). To achieve this, we insert a single-unit-cell-thick BFTO layer (ca.\ 4 nm) between BFO and LSMO. HAADF-STEM analysis of the resulting BFO/BFTO/LSMO/NGO heterostructure (Fig.\ \ref{fig:4}a) reveals an atomically sharp interface and the presence of a continuous Bi\textsubscript{2}O\textsubscript{2} interfacial layer beneath the BFO film. The heterostructure maintains coherent strain from the NGO (001) substrate, and thus BFO remains subject to the same anisotropic elastic boundary conditions as when grown directly on LSMO.

The interfacial Bi\textsubscript{2}O\textsubscript{2} layer at the BFTO/BFO interface consists of two Bi-O atomic planes (perovskite \textit{A}O planes) and a highly charged 2O\textsuperscript{4−} layer between them (Fig.\ \ref{fig:4}b). Since BFTO is insulating, these negative interfacial charges remain unscreened, generating a strong internal field, mimicking the previously seen effect of the ``skin layer'' and enforcing a uniform downward polarization in the BFO film. VPFM measurements corroborate this, showing a uniform out-of-plane response in pristine BFO on BFTO (1 u.c.)/LSMO (Fig.\ref{fig:4}c), unlike the nanoscale multidomain state in the LSMO-buffered sample (Fig.\ref{fig:1}c). Upon local switching, upward-oriented polarization components can be stabilized and subsequently reversed back to the original downward-oriented state, as evidenced by the two distinct VPFM contrast levels across the poled regions in Figure \ref{fig:4}c. This confirms that a controlled Bi\textsubscript{2}O\textsubscript{2} interfacial layer offers an effective strategy for stabilizing a uniform out-of-plane polarization in pristine BFO films, even under the strong anisotropic compressive strain imposed by NGO (001).

An LPFM measurement of the BFO film grown on the polarizing Bi\textsubscript{2}O\textsubscript{2} surface of the BFTO buffer (Fig.\ \ref{fig:4}d) reveals two in-plane domain variants in the pristine film, which, considering the uniform downward polarization and preserved uniaxial in-plane anisotropy (Fig.\ S2, Supporting Information), correspond to the $\vec{P}_1$ and $\vec{P}_2$ states. The domain walls between $\vec{P}_1$ and $\vec{P}_2$ are charged; here, however, the Bi\textsubscript{2}O\textsubscript{2} interfacial layer pins the net polarization direction downward in the pristine state and prevents local out-of-plane polarization reversals. But when the BFO out-of-plane polarization is poled upward with the scanning-probe tip, the newly formed CDWs between $\vec{P}_3$ and $\vec{P}_4$ are no longer stable. This is evidenced by local polarization back switching to the original downward-oriented polarization, emerging at the contours of the CDW (marked with an arrow in Fig.~\ref{fig:4}c). Note that comparable back-switching events were also observed in BFO grown directly on LSMO (cf.\ Fig.~\ref{fig:2}c). This indicates that once the out-of-plane polarization in BFO grown on BFTO no longer matches the screening charge provided by the Bi\textsubscript{2}O\textsubscript{2} interlayer, the strain-electrostatics frustration observed in the case of BFO/LSMO reappears. This demonstrates that the Bi\textsubscript{2}O\textsubscript{2} interlayer tunes the interfacial electrostatics while leaving the strain state unchanged, yet this single adjustment transforms the domain structure and domain-wall configurations in BFO.

\subsection{Signatures of chiral polar textures in the ferroelectric domain configurations}

Closer inspection of the LPFM image (Fig.~\ref{fig:4}d) reveals an unusual, yet systematic, lateral domain arrangement around the regions where the out-of-plane polarization has been locally poled upward. Outside the poled square, the pristine film shows a dark LPFM contrast on the left side, indicating only the $\vec{P}_1$ polarization variant, and a bright contrast on the right side, corresponding solely to $\vec{P}_2$. Similarly, within the upward-switched region, a distinct lateral organization of domain variants emerges: $\vec{P}_4$ tends to appear on the left-hand side, and $\vec{P}_3$ on the right-hand side. Lastly, in the reswitched area (smaller square), the $\vec{P}_2$  component selectively appears on the left-hand side and $\vec{P}_1$ on the right-hand side. This preferential arrangement of in-plane polarization variants around the switched and reswitched areas is a robust feature consistently found across many experiments on BFO films of various thicknesses grown on Bi\textsubscript{2}O\textsubscript{2}-terminated Aurivillius buffer layers, with the effect being most pronounced for thin BFO films where the interfacial influence is maximized. 

The deterministic domain sequence $\vec{P}_1$–$\vec{P}_4$–$\vec{P}_3$–$\vec{P}_2$–$\vec{P}_1$ (shown in the zoom-in of the LPFM scan, Fig.~\ref{fig:4}e, and schematized in Fig.~\ref{fig:4}f), is reminiscent of the uniform polarization rotation sense across all BFO domain walls and the 251\textdegree{} domain-wall type previously identified in BFO/BFTO flux-closure-like heterostructures\cite {gradauskaite2023}.  Across a 251\textdegree{} domain wall, the polarization is rotating between $\vec{P}_1$ and $\vec{P}_2$ components not through the shortest path via the downward component, but instead through the upward polarization (cf.\ Fig.\ \ref{fig:4}h). A similar configuration appears here: artificially upward-poled domains in our films consistently appear between $\vec{P}_1$ on their left and $\vec{P}_2$ on their right. Each switching event thus reinstates the deterministic domain sequence, so that the rotation of polarization components laterally across domains remains homochiral. This observation is consistent with a Dzyaloshinskii–Moriya–like interaction in ferroelectrics\cite{Zhao2021}, which energetically favors a particular polarization rotation direction in the film\cite{gradauskaite2023}.

In addition to the observed appearance of deterministic domain configurations, the VPFM images of our BFO films on BFTO also reveal discontinuities in the downward VPFM signal at domain walls, both in pristine films (see Fig.\ S3, Supporting Information) and in reswitched regions (Fig.~\ref{fig:4}g). These discontinuities indicate the changes in the out-of-plane polarization component upon crossing the wall. The downward-pointing component appears for all head-to-head domain walls, while the upward-pointing component is associated with tail-to-tail domain walls, as illustrated in Figure~\ref{fig:4}h, which confirms the presence of homochiral BFO domain walls\cite {gradauskaite2023} in our films as well. Going beyond the previous report on domain-wall homochirality in BFO films grown on Aurivillius buffers\cite{gradauskaite2023}, with this work, we demonstrate that homochiral domain-wall textures are preserved even after switching and re-switching the polarization back to its original downward state: domains expand upon poling, but the homochirality of the CDWs remains unchanged. We therefore observe polar homochirality in our films at two different length scales: in the deterministic domain configurations that appear upon poling, as well as the Néel-type domain walls themselves.

\begin{figure}[htb!]
  \centering
  \includegraphics[width=0.65\linewidth, clip, trim=4 7 4 4]{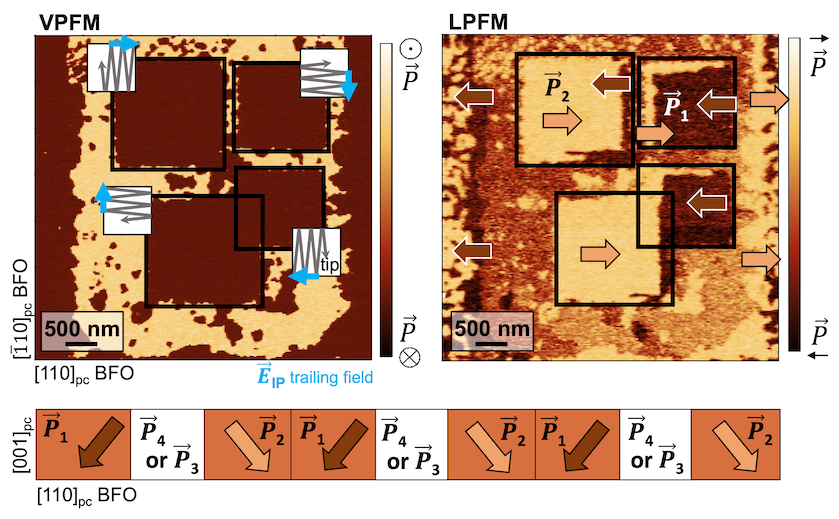}
  \caption{\textbf{Domain configuration reflecting polar homochirality in BFO film following polarization poling with differently oriented in-plane trailing fields.}  
  VPFM and LPFM images of a uniformly poled upward region in BFO/BFTO, where four square regions have been selectively reswitched downward. Although the in-plane trailing field applied during writing is rotated for each square (blue arrows, set by the slow-scan direction), every reswitched area exhibits the same chiral domain sequence: the $\vec{P}_1$ variant always develops on the left edge of an up-poled domain and the $\vec{P}_2$ variant on the right. Notably, this chirality is also evident at the edges of the larger upward-poled square, demonstrating the robustness of the polarization rotation pattern. The bottom schematic illustrates the domain sequence across the poled and reswitched areas in the cross-section of $(\overline{1}10)_{\mathrm{pc}}$ BFO plane, highlighting the preserved sense of polarization rotation previously observed at the 251\textdegree{} domain walls in BFO/BFTO\cite{gradauskaite2023}.}
  \label{fig:5}
\end{figure}

To further confirm that the observed domain sequences induced upon local poling are not coincidental, we artificially poled a large region upward and then selectively reversed the polarization in four square areas using different in-plane trailing field directions determined by the slow scanning axis (Fig.~\ref{fig:5}). In all four cases, the same domain sequence is preserved, with the $\vec{P}_2$ variant consistently appearing to the right of the upward-switched region and the $\vec{P}_1$ variant to the left. This effect extends beyond the four smaller reswitched squares, as the surrounding domains on the left and right sides of the large upward switched square follow the same chirality. While the trailing field influences the relative population of $\vec{P}_1$ and $\vec{P}_2$ within the switched squares, it does not alter the overall domain sequence, reinforcing the idea that the system exhibits a strong preference for homochirality. This persistence of the defined rotation of polarization components across the domain pattern suggests that the Dzyaloshinskii–Moriya–like interaction governs the polar architecture of the film both at the domain and domain-wall level.

\section{Conclusions}

Our work demonstrates that the polarization state in BFO thin films on LSMO-buffered NGO (001) arises from the interplay between anisotropic compressive strain, which promotes uniaxial in-plane polar anisotropy, and interfacial electrostatic boundary conditions that favor a uniform out-of-plane polarization. Although the charged atomic-plane termination of the LSMO buffer layer energetically favors a downward polarization, BFO instead adopts a multidomain state to minimize the electrostatic cost associated with CDWs. The spontaneous formation of a Bi\(_2\)O\(_2\) surface reconstruction in upward-polarized regions locally counteracts the electrostatic constraints imposed by the LSMO buffer layer. To harness this polarization compensation mechanism, we introduced an insulating Aurivillius buffer layer that establishes a uniform Bi\(_2\)O\(_2\) interfacial layer beneath the BFO film under the same anisotropic strain conditions. This approach allowed us to stabilize a uniform out-of-plane polarization with larger ferroelectric domains as well as CDWs. Moreover, we show a homochiral polarization rotation observable both at the level of polarization-component variation across domain configurations as well as at the non-Ising domain walls themselves that persist after polarization poling. These findings establish charged polarizing surfaces as a powerful tool for tuning domain configurations and polarization textures in oxide ferroelectrics.

\section{Experimental Section}
\RaggedRight

\subsection*{Thin-Film Growth}\justifying 
The thin films and heterostructures were grown on NGO (001) substrates by pulsed laser deposition using a KrF excimer laser at 248 nm. The laser fluence, repetition rate, substrate temperature, and growth pressure set for individual layers are as follows: LSMO: 0.9 J cm\textsuperscript{−2}, 1 Hz, 650 \textdegree{}C, 0.035 mbar O\textsubscript{2}, BFO: 0.7 J cm\textsuperscript{−2}, 8 Hz, 630 \textdegree{}C, 0.125 mbar O\textsubscript{2}; BFTO: 0.9 J cm\textsuperscript{−2}, 2 Hz, 650 \textdegree{}C, 0.075 mbar O\textsubscript{2}. The thickness of the thin films was monitored using a combination of RHEED during growth and X-ray reflectivity ex-situ.

\subsection*{X-ray Diffraction}\justifying  The crystalline structure of thin films was analyzed by X-ray diffraction, X-ray reflectivity, and reciprocal space mapping measurements using a four-circle X-ray diffractometer (Panalytical X'Pert MRD).

\subsection*{Scanning Probe Microscopy}\justifying 
The scanning probe microscopy measurements were performed using both an NT-MDT NTEGRA scanning-probe microscope and a Bruker Nanoscope V Multimode system with two external SR830 lock-in detectors (Stanford Research) for simultaneous acquisition of in-plane and out-of-plane piezoresponse. Measurements on the NT-MDT system were carried out using a 2.5-V peak-to-peak AC modulation at 70 kHz applied to a Pt-coated tip (HQ:NSC35/Pt), with deflection and torsion modes recorded when the cantilever was positioned perpendicular to the uniaxial polarization axis of BFO ([110]\textsubscript{pc}). Additional PFM experiments were conducted using the Bruker Nanoscope V Multimode microscope. Here, AC modulation voltages of 2.5 V at 35 kHz were applied to a Pt-coated tip (BudgetSensors, 40 N/m cantilever). Out-of-plane polarization was locally poled using a 7-8 V DC voltage applied to the tip.

 \subsection*{Scanning Transmission Electron Microscopy (STEM)}\justifying  Cross-sectional specimens were prepared by focused ion beam (FIB) milling with a FEI Helios NanoLab 600i. High-angle annular dark-field scanning transmission electron microscopy (HAADF-STEM) imaging was achieved using a FEI Titan Themis microscope equipped with a spherical-aberration probe corrector (CEOS DCOR), operated at 300 kV. A probe semi-convergence angle of 18 mrad was set in combination with an annular semi-detection range of the HAADF detector of 66-200 mrad. The HAADF-STEM images were obtained by averaging time series of 10 frames, and the atomic column positions in the averaged images were fitted with picometer precision using custom-developed scripts, as described by Campanini \textit{et al.}\cite{Campanini2018}. Ferroelectric dipole maps were calculated from the relative displacements of the two cation sublattices present in the perovskite-type structures with the general formula of \textit{AB}O\textsubscript{3}. Finally, the polarization map derived from the HAADF-STEM image in Figure \ref{fig:2}e is shown with the polarization vectors opposite to the displacement of the $B$ cations (Fig.\ \ref{fig:2}f).

\section*{Acknowledgements}
E.G. and M.T. acknowledge the Swiss National Science Foundation under Project No. 200021\_188414. E.G. acknowledges the Swiss National Science Foundation for financial support under Project No.\ P500PT\_214449.  M.T. acknowledges the Swiss National Science Foundation under Project No. 200021\_231428. M.D.R. acknowledges the Swiss National Science Foundation under Project No.
200021\_175926.

\section*{Conflict of Interest} The authors declare no conflict of interest.

\section*{Author contributions}
E.G. performed the thin-film growth, scanning-probe microscopy, and structural thin-film analysis. M.D.R. carried out the STEM investigations. N.G. assisted with the piezoresponse force microscopy measurements. Q.N.M. contributed to identifying and interpreting the homochiral polarization rotation patterns. E.G. designed the experiment. M.T. supervised the work. The manuscript was written by E.G. and M.T. with input from all authors.


\bibliographystyle{apsrev4-2}
\bibliography{frustrated_BFO}

\begin{thebibliography}{43}%
\makeatletter
\providecommand \@ifxundefined [1]{%
 \@ifx{#1\undefined}
}%
\providecommand \@ifnum [1]{%
 \ifnum #1\expandafter \@firstoftwo
 \else \expandafter \@secondoftwo
 \fi
}%
\providecommand \@ifx [1]{%
 \ifx #1\expandafter \@firstoftwo
 \else \expandafter \@secondoftwo
 \fi
}%
\providecommand \natexlab [1]{#1}%
\providecommand \enquote  [1]{``#1''}%
\providecommand \bibnamefont  [1]{#1}%
\providecommand \bibfnamefont [1]{#1}%
\providecommand \citenamefont [1]{#1}%
\providecommand \href@noop [0]{\@secondoftwo}%
\providecommand \href [0]{\begingroup \@sanitize@url \@href}%
\providecommand \@href[1]{\@@startlink{#1}\@@href}%
\providecommand \@@href[1]{\endgroup#1\@@endlink}%
\providecommand \@sanitize@url [0]{\catcode `\\12\catcode `\$12\catcode `\&12\catcode `\#12\catcode `\^12\catcode `\_12\catcode `\%12\relax}%
\providecommand \@@startlink[1]{}%
\providecommand \@@endlink[0]{}%
\providecommand \url  [0]{\begingroup\@sanitize@url \@url }%
\providecommand \@url [1]{\endgroup\@href {#1}{\urlprefix }}%
\providecommand \urlprefix  [0]{URL }%
\providecommand \Eprint [0]{\href }%
\providecommand \doibase [0]{https://doi.org/}%
\providecommand \selectlanguage [0]{\@gobble}%
\providecommand \bibinfo  [0]{\@secondoftwo}%
\providecommand \bibfield  [0]{\@secondoftwo}%
\providecommand \translation [1]{[#1]}%
\providecommand \BibitemOpen [0]{}%
\providecommand \bibitemStop [0]{}%
\providecommand \bibitemNoStop [0]{.\EOS\space}%
\providecommand \EOS [0]{\spacefactor3000\relax}%
\providecommand \BibitemShut  [1]{\csname bibitem#1\endcsname}%
\let\auto@bib@innerbib\@empty
\bibitem [{\citenamefont {Scott}(2000)}]{Scott2000book}%
  \BibitemOpen
  \bibfield  {author} {\bibinfo {author} {\bibfnamefont {J.~F.}\ \bibnamefont {Scott}},\ }\href@noop {} {\emph {\bibinfo {title} {Ferroelectric {{Memories}}}}}\ (\bibinfo {address} {New York},\ \bibinfo {year} {2000})\BibitemShut {NoStop}%
\bibitem [{\citenamefont {Müller}\ \emph {et~al.}(2023)\citenamefont {Müller}, \citenamefont {Efe}, \citenamefont {Sarott}, \citenamefont {Gradauskaite},\ and\ \citenamefont {Trassin}}]{muller2023}%
  \BibitemOpen
  \bibfield  {author} {\bibinfo {author} {\bibfnamefont {M.}~\bibnamefont {Müller}}, \bibinfo {author} {\bibfnamefont {I.}~\bibnamefont {Efe}}, \bibinfo {author} {\bibfnamefont {M.~F.}\ \bibnamefont {Sarott}}, \bibinfo {author} {\bibfnamefont {E.}~\bibnamefont {Gradauskaite}},\ and\ \bibinfo {author} {\bibfnamefont {M.}~\bibnamefont {Trassin}},\ }\href {https://doi.org/10.1021/acsaelm.2c01755} {\bibfield  {journal} {\bibinfo  {journal} {ACS Applied Electronic Materials}\ }\textbf {\bibinfo {volume} {5}},\ \bibinfo {pages} {1314} (\bibinfo {year} {2023})}\BibitemShut {NoStop}%
\bibitem [{\citenamefont {Lichtensteiger}\ \emph {et~al.}(2014)\citenamefont {Lichtensteiger}, \citenamefont {{Fernandez-Pena}}, \citenamefont {Weymann}, \citenamefont {Zubko},\ and\ \citenamefont {Triscone}}]{Lichtensteiger2014a}%
  \BibitemOpen
  \bibfield  {author} {\bibinfo {author} {\bibfnamefont {C.}~\bibnamefont {Lichtensteiger}}, \bibinfo {author} {\bibfnamefont {S.}~\bibnamefont {{Fernandez-Pena}}}, \bibinfo {author} {\bibfnamefont {C.}~\bibnamefont {Weymann}}, \bibinfo {author} {\bibfnamefont {P.}~\bibnamefont {Zubko}},\ and\ \bibinfo {author} {\bibfnamefont {J.~M.}\ \bibnamefont {Triscone}},\ }\href {https://doi.org/10.1021/nl404734z} {\bibfield  {journal} {\bibinfo  {journal} {Nano Letters}\ }\textbf {\bibinfo {volume} {14}},\ \bibinfo {pages} {4205} (\bibinfo {year} {2014})}\BibitemShut {NoStop}%
\bibitem [{\citenamefont {Strkalj}\ \emph {et~al.}(2019)\citenamefont {Strkalj}, \citenamefont {De~Luca}, \citenamefont {Campanini}, \citenamefont {Pal}, \citenamefont {Schaab}, \citenamefont {Gattinoni}, \citenamefont {Spaldin}, \citenamefont {Rossell}, \citenamefont {Fiebig},\ and\ \citenamefont {Trassin}}]{Strkalj2019}%
  \BibitemOpen
  \bibfield  {author} {\bibinfo {author} {\bibfnamefont {N.}~\bibnamefont {Strkalj}}, \bibinfo {author} {\bibfnamefont {G.}~\bibnamefont {De~Luca}}, \bibinfo {author} {\bibfnamefont {M.}~\bibnamefont {Campanini}}, \bibinfo {author} {\bibfnamefont {S.}~\bibnamefont {Pal}}, \bibinfo {author} {\bibfnamefont {J.}~\bibnamefont {Schaab}}, \bibinfo {author} {\bibfnamefont {C.}~\bibnamefont {Gattinoni}}, \bibinfo {author} {\bibfnamefont {N.~A.}\ \bibnamefont {Spaldin}}, \bibinfo {author} {\bibfnamefont {M.~D.}\ \bibnamefont {Rossell}}, \bibinfo {author} {\bibfnamefont {M.}~\bibnamefont {Fiebig}},\ and\ \bibinfo {author} {\bibfnamefont {M.}~\bibnamefont {Trassin}},\ }\href {https://doi.org/10.1103/PhysRevLett.123.147601} {\bibfield  {journal} {\bibinfo  {journal} {Physical Review Letters}\ }\textbf {\bibinfo {volume} {123}},\ \bibinfo {pages} {147601} (\bibinfo {year} {2019})}\BibitemShut {NoStop}%
\bibitem [{\citenamefont {Strkalj}\ \emph {et~al.}(2020)\citenamefont {Strkalj}, \citenamefont {Gattinoni}, \citenamefont {Vogel}, \citenamefont {Campanini}, \citenamefont {Haerdi}, \citenamefont {Rossi}, \citenamefont {Rossell}, \citenamefont {Spaldin}, \citenamefont {Fiebig},\ and\ \citenamefont {Trassin}}]{Strkalj2020}%
  \BibitemOpen
  \bibfield  {author} {\bibinfo {author} {\bibfnamefont {N.}~\bibnamefont {Strkalj}}, \bibinfo {author} {\bibfnamefont {C.}~\bibnamefont {Gattinoni}}, \bibinfo {author} {\bibfnamefont {A.}~\bibnamefont {Vogel}}, \bibinfo {author} {\bibfnamefont {M.}~\bibnamefont {Campanini}}, \bibinfo {author} {\bibfnamefont {R.}~\bibnamefont {Haerdi}}, \bibinfo {author} {\bibfnamefont {A.}~\bibnamefont {Rossi}}, \bibinfo {author} {\bibfnamefont {M.~D.}\ \bibnamefont {Rossell}}, \bibinfo {author} {\bibfnamefont {N.~A.}\ \bibnamefont {Spaldin}}, \bibinfo {author} {\bibfnamefont {M.}~\bibnamefont {Fiebig}},\ and\ \bibinfo {author} {\bibfnamefont {M.}~\bibnamefont {Trassin}},\ }\href {https://doi.org/10.1038/s41467-020-19635-7} {\bibfield  {journal} {\bibinfo  {journal} {Nature Communications}\ }\textbf {\bibinfo {volume} {11}},\ \bibinfo {pages} {5815} (\bibinfo {year} {2020})}\BibitemShut {NoStop}%
\bibitem [{\citenamefont {Efe}\ \emph {et~al.}(2024)\citenamefont {Efe}, \citenamefont {Yan},\ and\ \citenamefont {Trassin}}]{efe2024}%
  \BibitemOpen
  \bibfield  {author} {\bibinfo {author} {\bibfnamefont {I.}~\bibnamefont {Efe}}, \bibinfo {author} {\bibfnamefont {B.}~\bibnamefont {Yan}},\ and\ \bibinfo {author} {\bibfnamefont {M.}~\bibnamefont {Trassin}},\ }\href {https://doi.org/10.1063/5.0232382} {\bibfield  {journal} {\bibinfo  {journal} {Applied Physics Letters}\ }\textbf {\bibinfo {volume} {125}},\ \bibinfo {pages} {150503} (\bibinfo {year} {2024})}\BibitemShut {NoStop}%
\bibitem [{\citenamefont {Zeches}\ \emph {et~al.}(2009)\citenamefont {Zeches}, \citenamefont {Rossell}, \citenamefont {Zhang}, \citenamefont {Hatt}, \citenamefont {He}, \citenamefont {Yang}, \citenamefont {Kumar}, \citenamefont {Wang}, \citenamefont {Melville}, \citenamefont {Adamo}, \citenamefont {Sheng}, \citenamefont {Chu}, \citenamefont {Ihlefeld}, \citenamefont {Erni}, \citenamefont {Ederer}, \citenamefont {Gopalan}, \citenamefont {Chen}, \citenamefont {Schlom}, \citenamefont {Spaldin}, \citenamefont {Martin},\ and\ \citenamefont {Ramesh}}]{Zeches2009}%
  \BibitemOpen
  \bibfield  {author} {\bibinfo {author} {\bibfnamefont {R.~J.}\ \bibnamefont {Zeches}}, \bibinfo {author} {\bibfnamefont {M.~D.}\ \bibnamefont {Rossell}}, \bibinfo {author} {\bibfnamefont {J.~X.}\ \bibnamefont {Zhang}}, \bibinfo {author} {\bibfnamefont {A.~J.}\ \bibnamefont {Hatt}}, \bibinfo {author} {\bibfnamefont {Q.}~\bibnamefont {He}}, \bibinfo {author} {\bibfnamefont {C.-H.}\ \bibnamefont {Yang}}, \bibinfo {author} {\bibfnamefont {A.}~\bibnamefont {Kumar}}, \bibinfo {author} {\bibfnamefont {C.~H.}\ \bibnamefont {Wang}}, \bibinfo {author} {\bibfnamefont {A.}~\bibnamefont {Melville}}, \bibinfo {author} {\bibfnamefont {C.}~\bibnamefont {Adamo}}, \bibinfo {author} {\bibfnamefont {G.}~\bibnamefont {Sheng}}, \bibinfo {author} {\bibfnamefont {Y.-H.}\ \bibnamefont {Chu}}, \bibinfo {author} {\bibfnamefont {J.~F.}\ \bibnamefont {Ihlefeld}}, \bibinfo {author} {\bibfnamefont {R.}~\bibnamefont {Erni}}, \bibinfo {author} {\bibfnamefont {C.}~\bibnamefont {Ederer}}, \bibinfo {author} {\bibfnamefont {V.}~\bibnamefont
  {Gopalan}}, \bibinfo {author} {\bibfnamefont {L.~Q.}\ \bibnamefont {Chen}}, \bibinfo {author} {\bibfnamefont {D.~G.}\ \bibnamefont {Schlom}}, \bibinfo {author} {\bibfnamefont {N.~A.}\ \bibnamefont {Spaldin}}, \bibinfo {author} {\bibfnamefont {L.~W.}\ \bibnamefont {Martin}},\ and\ \bibinfo {author} {\bibfnamefont {R.}~\bibnamefont {Ramesh}},\ }\href {https://doi.org/10.1126/science.1177046} {\bibfield  {journal} {\bibinfo  {journal} {Science}\ }\textbf {\bibinfo {volume} {326}},\ \bibinfo {pages} {977} (\bibinfo {year} {2009})}\BibitemShut {NoStop}%
\bibitem [{\citenamefont {Damodaran}\ \emph {et~al.}(2014)\citenamefont {Damodaran}, \citenamefont {Breckenfeld}, \citenamefont {Chen}, \citenamefont {Lee},\ and\ \citenamefont {Martin}}]{Damodaran2014}%
  \BibitemOpen
  \bibfield  {author} {\bibinfo {author} {\bibfnamefont {A.~R.}\ \bibnamefont {Damodaran}}, \bibinfo {author} {\bibfnamefont {E.}~\bibnamefont {Breckenfeld}}, \bibinfo {author} {\bibfnamefont {Z.}~\bibnamefont {Chen}}, \bibinfo {author} {\bibfnamefont {S.}~\bibnamefont {Lee}},\ and\ \bibinfo {author} {\bibfnamefont {L.~W.}\ \bibnamefont {Martin}},\ }\href {https://doi.org/10.1002/adma.201400254} {\bibfield  {journal} {\bibinfo  {journal} {Advanced Materials}\ }\textbf {\bibinfo {volume} {26}},\ \bibinfo {pages} {6341} (\bibinfo {year} {2014})}\BibitemShut {NoStop}%
\bibitem [{\citenamefont {Nordlander}\ \emph {et~al.}(2024)\citenamefont {Nordlander}, \citenamefont {Grosso}, \citenamefont {Rossell}, \citenamefont {Maillard}, \citenamefont {Yan}, \citenamefont {Gradauskaite}, \citenamefont {Spaldin}, \citenamefont {Fiebig},\ and\ \citenamefont {Trassin}}]{nordlander2024}%
  \BibitemOpen
  \bibfield  {author} {\bibinfo {author} {\bibfnamefont {J.}~\bibnamefont {Nordlander}}, \bibinfo {author} {\bibfnamefont {B.~F.}\ \bibnamefont {Grosso}}, \bibinfo {author} {\bibfnamefont {M.~D.}\ \bibnamefont {Rossell}}, \bibinfo {author} {\bibfnamefont {A.}~\bibnamefont {Maillard}}, \bibinfo {author} {\bibfnamefont {B.}~\bibnamefont {Yan}}, \bibinfo {author} {\bibfnamefont {E.}~\bibnamefont {Gradauskaite}}, \bibinfo {author} {\bibfnamefont {N.~A.}\ \bibnamefont {Spaldin}}, \bibinfo {author} {\bibfnamefont {M.}~\bibnamefont {Fiebig}},\ and\ \bibinfo {author} {\bibfnamefont {M.}~\bibnamefont {Trassin}},\ }\href {https://doi.org/10.1002/aelm.202400185} {\bibfield  {journal} {\bibinfo  {journal} {Advanced Electronic Materials}\ }\textbf {\bibinfo {volume} {10}},\ \bibinfo {pages} {2400185} (\bibinfo {year} {2024})}\BibitemShut {NoStop}%
\bibitem [{\citenamefont {Li}\ \emph {et~al.}(2018)\citenamefont {Li}, \citenamefont {Jokisaari}, \citenamefont {Zhang}, \citenamefont {Cheng}, \citenamefont {Yan}, \citenamefont {Heikes}, \citenamefont {Lin}, \citenamefont {Gadre}, \citenamefont {Schlom}, \citenamefont {Chen},\ and\ \citenamefont {Pan}}]{Li2018a}%
  \BibitemOpen
  \bibfield  {author} {\bibinfo {author} {\bibfnamefont {L.}~\bibnamefont {Li}}, \bibinfo {author} {\bibfnamefont {J.~R.}\ \bibnamefont {Jokisaari}}, \bibinfo {author} {\bibfnamefont {Y.}~\bibnamefont {Zhang}}, \bibinfo {author} {\bibfnamefont {X.}~\bibnamefont {Cheng}}, \bibinfo {author} {\bibfnamefont {X.}~\bibnamefont {Yan}}, \bibinfo {author} {\bibfnamefont {C.}~\bibnamefont {Heikes}}, \bibinfo {author} {\bibfnamefont {Q.}~\bibnamefont {Lin}}, \bibinfo {author} {\bibfnamefont {C.}~\bibnamefont {Gadre}}, \bibinfo {author} {\bibfnamefont {D.~G.}\ \bibnamefont {Schlom}}, \bibinfo {author} {\bibfnamefont {L.~Q.}\ \bibnamefont {Chen}},\ and\ \bibinfo {author} {\bibfnamefont {X.}~\bibnamefont {Pan}},\ }\href {https://doi.org/10.1002/adma.201802737} {\bibfield  {journal} {\bibinfo  {journal} {Advanced Materials}\ }\textbf {\bibinfo {volume} {30}},\ \bibinfo {pages} {1802737} (\bibinfo {year} {2018})}\BibitemShut {NoStop}%
\bibitem [{\citenamefont {Weymann}\ \emph {et~al.}(2020)\citenamefont {Weymann}, \citenamefont {Lichtensteiger}, \citenamefont {{Fernandez-Peña}}, \citenamefont {Naden}, \citenamefont {Dedon}, \citenamefont {Martin}, \citenamefont {Triscone},\ and\ \citenamefont {Paruch}}]{Weymann2020}%
  \BibitemOpen
  \bibfield  {author} {\bibinfo {author} {\bibfnamefont {C.}~\bibnamefont {Weymann}}, \bibinfo {author} {\bibfnamefont {C.}~\bibnamefont {Lichtensteiger}}, \bibinfo {author} {\bibfnamefont {S.}~\bibnamefont {{Fernandez-Peña}}}, \bibinfo {author} {\bibfnamefont {A.~B.}\ \bibnamefont {Naden}}, \bibinfo {author} {\bibfnamefont {L.~R.}\ \bibnamefont {Dedon}}, \bibinfo {author} {\bibfnamefont {L.~W.}\ \bibnamefont {Martin}}, \bibinfo {author} {\bibfnamefont {J.~M.}\ \bibnamefont {Triscone}},\ and\ \bibinfo {author} {\bibfnamefont {P.}~\bibnamefont {Paruch}},\ }\href {https://doi.org/10.1002/aelm.202000852} {\bibfield  {journal} {\bibinfo  {journal} {Advanced Electronic Materials}\ }\textbf {\bibinfo {volume} {6}},\ \bibinfo {pages} {2000852} (\bibinfo {year} {2020})}\BibitemShut {NoStop}%
\bibitem [{\citenamefont {Gradauskaite}\ \emph {et~al.}(2022)\citenamefont {Gradauskaite}, \citenamefont {Hunnestad}, \citenamefont {Meier}, \citenamefont {Meier},\ and\ \citenamefont {Trassin}}]{gradauskaite2022}%
  \BibitemOpen
  \bibfield  {author} {\bibinfo {author} {\bibfnamefont {E.}~\bibnamefont {Gradauskaite}}, \bibinfo {author} {\bibfnamefont {K.~A.}\ \bibnamefont {Hunnestad}}, \bibinfo {author} {\bibfnamefont {Q.~N.}\ \bibnamefont {Meier}}, \bibinfo {author} {\bibfnamefont {D.}~\bibnamefont {Meier}},\ and\ \bibinfo {author} {\bibfnamefont {M.}~\bibnamefont {Trassin}},\ }\href {https://doi.org/10.1021/acs.chemmater.2c01178} {\bibfield  {journal} {\bibinfo  {journal} {Chemistry of Materials}\ }\textbf {\bibinfo {volume} {34}},\ \bibinfo {pages} {6468} (\bibinfo {year} {2022})}\BibitemShut {NoStop}%
\bibitem [{\citenamefont {Sarott}\ \emph {et~al.}(2023)\citenamefont {Sarott}, \citenamefont {Bucheli}, \citenamefont {Lochmann}, \citenamefont {Fiebig},\ and\ \citenamefont {Trassin}}]{sarott2023}%
  \BibitemOpen
  \bibfield  {author} {\bibinfo {author} {\bibfnamefont {M.~F.}\ \bibnamefont {Sarott}}, \bibinfo {author} {\bibfnamefont {U.}~\bibnamefont {Bucheli}}, \bibinfo {author} {\bibfnamefont {A.}~\bibnamefont {Lochmann}}, \bibinfo {author} {\bibfnamefont {M.}~\bibnamefont {Fiebig}},\ and\ \bibinfo {author} {\bibfnamefont {M.}~\bibnamefont {Trassin}},\ }\href {https://doi.org/10.1002/adfm.202214849} {\bibfield  {journal} {\bibinfo  {journal} {Advanced Functional Materials}\ }\textbf {\bibinfo {volume} {33}},\ \bibinfo {pages} {2214849} (\bibinfo {year} {2023})}\BibitemShut {NoStop}%
\bibitem [{\citenamefont {Yu}\ \emph {et~al.}(2012)\citenamefont {Yu}, \citenamefont {Luo}, \citenamefont {Yi}, \citenamefont {Zhang}, \citenamefont {Rossell}, \citenamefont {Yang}, \citenamefont {You}, \citenamefont {{Singh-Bhalla}}, \citenamefont {Yang}, \citenamefont {He}, \citenamefont {Ramasse}, \citenamefont {Erni}, \citenamefont {Martin}, \citenamefont {Chu}, \citenamefont {Pantelides}, \citenamefont {Pennycook},\ and\ \citenamefont {Ramesh}}]{Yu2012a}%
  \BibitemOpen
  \bibfield  {author} {\bibinfo {author} {\bibfnamefont {P.}~\bibnamefont {Yu}}, \bibinfo {author} {\bibfnamefont {W.}~\bibnamefont {Luo}}, \bibinfo {author} {\bibfnamefont {D.}~\bibnamefont {Yi}}, \bibinfo {author} {\bibfnamefont {J.~X.}\ \bibnamefont {Zhang}}, \bibinfo {author} {\bibfnamefont {M.~D.}\ \bibnamefont {Rossell}}, \bibinfo {author} {\bibfnamefont {C.-H.}\ \bibnamefont {Yang}}, \bibinfo {author} {\bibfnamefont {L.}~\bibnamefont {You}}, \bibinfo {author} {\bibfnamefont {G.}~\bibnamefont {{Singh-Bhalla}}}, \bibinfo {author} {\bibfnamefont {S.~Y.}\ \bibnamefont {Yang}}, \bibinfo {author} {\bibfnamefont {Q.}~\bibnamefont {He}}, \bibinfo {author} {\bibfnamefont {Q.~M.}\ \bibnamefont {Ramasse}}, \bibinfo {author} {\bibfnamefont {R.}~\bibnamefont {Erni}}, \bibinfo {author} {\bibfnamefont {L.~W.}\ \bibnamefont {Martin}}, \bibinfo {author} {\bibfnamefont {Y.~H.}\ \bibnamefont {Chu}}, \bibinfo {author} {\bibfnamefont {S.~T.}\ \bibnamefont {Pantelides}}, \bibinfo {author} {\bibfnamefont {S.~J.}\
  \bibnamefont {Pennycook}},\ and\ \bibinfo {author} {\bibfnamefont {R.}~\bibnamefont {Ramesh}},\ }\href {https://doi.org/10.1073/pnas.1117990109} {\bibfield  {journal} {\bibinfo  {journal} {Proceedings of the National Academy of Sciences}\ }\textbf {\bibinfo {volume} {109}},\ \bibinfo {pages} {9710} (\bibinfo {year} {2012})}\BibitemShut {NoStop}%
\bibitem [{\citenamefont {Junquera}\ and\ \citenamefont {Ghosez}(2003)}]{Junquera2003}%
  \BibitemOpen
  \bibfield  {author} {\bibinfo {author} {\bibfnamefont {J.}~\bibnamefont {Junquera}}\ and\ \bibinfo {author} {\bibfnamefont {P.}~\bibnamefont {Ghosez}},\ }\href {https://doi.org/10.1038/nature01501} {\bibfield  {journal} {\bibinfo  {journal} {Nature}\ }\textbf {\bibinfo {volume} {422}},\ \bibinfo {pages} {506} (\bibinfo {year} {2003})}\BibitemShut {NoStop}%
\bibitem [{\citenamefont {Kim}\ \emph {et~al.}(2005)\citenamefont {Kim}, \citenamefont {Kim}, \citenamefont {Kim}, \citenamefont {Chang}, \citenamefont {Noh}, \citenamefont {Kong}, \citenamefont {Char}, \citenamefont {Park}, \citenamefont {Bu}, \citenamefont {Yoon},\ and\ \citenamefont {Chung}}]{Kim2005a}%
  \BibitemOpen
  \bibfield  {author} {\bibinfo {author} {\bibfnamefont {Y.~S.}\ \bibnamefont {Kim}}, \bibinfo {author} {\bibfnamefont {D.~H.}\ \bibnamefont {Kim}}, \bibinfo {author} {\bibfnamefont {J.~D.}\ \bibnamefont {Kim}}, \bibinfo {author} {\bibfnamefont {Y.~J.}\ \bibnamefont {Chang}}, \bibinfo {author} {\bibfnamefont {T.~W.}\ \bibnamefont {Noh}}, \bibinfo {author} {\bibfnamefont {J.~H.}\ \bibnamefont {Kong}}, \bibinfo {author} {\bibfnamefont {K.}~\bibnamefont {Char}}, \bibinfo {author} {\bibfnamefont {Y.~D.}\ \bibnamefont {Park}}, \bibinfo {author} {\bibfnamefont {S.~D.}\ \bibnamefont {Bu}}, \bibinfo {author} {\bibfnamefont {J.~G.}\ \bibnamefont {Yoon}},\ and\ \bibinfo {author} {\bibfnamefont {J.~S.}\ \bibnamefont {Chung}},\ }\href {https://doi.org/10.1063/1.1880443} {\bibfield  {journal} {\bibinfo  {journal} {Applied Physics Letters}\ }\textbf {\bibinfo {volume} {86}},\ \bibinfo {pages} {102907} (\bibinfo {year} {2005})}\BibitemShut {NoStop}%
\bibitem [{\citenamefont {Schlom}\ \emph {et~al.}(2007)\citenamefont {Schlom}, \citenamefont {Chen}, \citenamefont {Eom}, \citenamefont {Rabe}, \citenamefont {Streiffer},\ and\ \citenamefont {Triscone}}]{Schlom2007}%
  \BibitemOpen
  \bibfield  {author} {\bibinfo {author} {\bibfnamefont {D.~G.}\ \bibnamefont {Schlom}}, \bibinfo {author} {\bibfnamefont {L.~Q.}\ \bibnamefont {Chen}}, \bibinfo {author} {\bibfnamefont {C.~B.}\ \bibnamefont {Eom}}, \bibinfo {author} {\bibfnamefont {K.~M.}\ \bibnamefont {Rabe}}, \bibinfo {author} {\bibfnamefont {S.~K.}\ \bibnamefont {Streiffer}},\ and\ \bibinfo {author} {\bibfnamefont {J.~M.}\ \bibnamefont {Triscone}},\ }\href {https://doi.org/10.1146/annurev.matsci.37.061206.113016} {\bibfield  {journal} {\bibinfo  {journal} {Annual Review of Materials Research}\ }\textbf {\bibinfo {volume} {37}},\ \bibinfo {pages} {589} (\bibinfo {year} {2007})}\BibitemShut {NoStop}%
\bibitem [{\citenamefont {Chen}\ \emph {et~al.}(2014)\citenamefont {Chen}, \citenamefont {Damodaran}, \citenamefont {Xu}, \citenamefont {Lee},\ and\ \citenamefont {Martin}}]{Chen2014}%
  \BibitemOpen
  \bibfield  {author} {\bibinfo {author} {\bibfnamefont {Z.~H.}\ \bibnamefont {Chen}}, \bibinfo {author} {\bibfnamefont {A.~R.}\ \bibnamefont {Damodaran}}, \bibinfo {author} {\bibfnamefont {R.}~\bibnamefont {Xu}}, \bibinfo {author} {\bibfnamefont {S.}~\bibnamefont {Lee}},\ and\ \bibinfo {author} {\bibfnamefont {L.~W.}\ \bibnamefont {Martin}},\ }\href {https://doi.org/10.1063/1.4875801} {\bibfield  {journal} {\bibinfo  {journal} {Applied Physics Letters}\ }\textbf {\bibinfo {volume} {104}},\ \bibinfo {pages} {182908} (\bibinfo {year} {2014})}\BibitemShut {NoStop}%
\bibitem [{\citenamefont {Catalan}\ \emph {et~al.}(2006)\citenamefont {Catalan}, \citenamefont {Janssens}, \citenamefont {Rispens}, \citenamefont {Csiszar}, \citenamefont {Seeck}, \citenamefont {Rijnders}, \citenamefont {Blank},\ and\ \citenamefont {Noheda}}]{catalan2006a}%
  \BibitemOpen
  \bibfield  {author} {\bibinfo {author} {\bibfnamefont {G.}~\bibnamefont {Catalan}}, \bibinfo {author} {\bibfnamefont {A.}~\bibnamefont {Janssens}}, \bibinfo {author} {\bibfnamefont {G.}~\bibnamefont {Rispens}}, \bibinfo {author} {\bibfnamefont {S.}~\bibnamefont {Csiszar}}, \bibinfo {author} {\bibfnamefont {O.}~\bibnamefont {Seeck}}, \bibinfo {author} {\bibfnamefont {G.}~\bibnamefont {Rijnders}}, \bibinfo {author} {\bibfnamefont {D.~H.~A.}\ \bibnamefont {Blank}},\ and\ \bibinfo {author} {\bibfnamefont {B.}~\bibnamefont {Noheda}},\ }\href {https://doi.org/10.1103/physrevlett.96.127602} {\bibfield  {journal} {\bibinfo  {journal} {Physical Review Letters}\ }\textbf {\bibinfo {volume} {96}},\ \bibinfo {pages} {127602} (\bibinfo {year} {2006})}\BibitemShut {NoStop}%
\bibitem [{\citenamefont {Becher}\ \emph {et~al.}(2015)\citenamefont {Becher}, \citenamefont {Maurel}, \citenamefont {Aschauer}, \citenamefont {Lilienblum}, \citenamefont {Magén}, \citenamefont {Meier}, \citenamefont {Langenberg}, \citenamefont {Trassin}, \citenamefont {Blasco}, \citenamefont {Krug}, \citenamefont {Algarabel}, \citenamefont {Spaldin}, \citenamefont {Pardo},\ and\ \citenamefont {Fiebig}}]{Becher2015a}%
  \BibitemOpen
  \bibfield  {author} {\bibinfo {author} {\bibfnamefont {C.}~\bibnamefont {Becher}}, \bibinfo {author} {\bibfnamefont {L.}~\bibnamefont {Maurel}}, \bibinfo {author} {\bibfnamefont {U.}~\bibnamefont {Aschauer}}, \bibinfo {author} {\bibfnamefont {M.}~\bibnamefont {Lilienblum}}, \bibinfo {author} {\bibfnamefont {C.}~\bibnamefont {Magén}}, \bibinfo {author} {\bibfnamefont {D.}~\bibnamefont {Meier}}, \bibinfo {author} {\bibfnamefont {E.}~\bibnamefont {Langenberg}}, \bibinfo {author} {\bibfnamefont {M.}~\bibnamefont {Trassin}}, \bibinfo {author} {\bibfnamefont {J.}~\bibnamefont {Blasco}}, \bibinfo {author} {\bibfnamefont {I.~P.}\ \bibnamefont {Krug}}, \bibinfo {author} {\bibfnamefont {P.~A.}\ \bibnamefont {Algarabel}}, \bibinfo {author} {\bibfnamefont {N.~A.}\ \bibnamefont {Spaldin}}, \bibinfo {author} {\bibfnamefont {J.~A.}\ \bibnamefont {Pardo}},\ and\ \bibinfo {author} {\bibfnamefont {M.}~\bibnamefont {Fiebig}},\ }\href {https://doi.org/10.1038/nnano.2015.108} {\bibfield  {journal} {\bibinfo  {journal} {Nature
  Nanotechnology}\ }\textbf {\bibinfo {volume} {10}},\ \bibinfo {pages} {661} (\bibinfo {year} {2015})}\BibitemShut {NoStop}%
\bibitem [{\citenamefont {Chu}\ \emph {et~al.}(2009)\citenamefont {Chu}, \citenamefont {He}, \citenamefont {Yang}, \citenamefont {Yu}, \citenamefont {Martin}, \citenamefont {Shafer},\ and\ \citenamefont {Ramesh}}]{Chu2009}%
  \BibitemOpen
  \bibfield  {author} {\bibinfo {author} {\bibfnamefont {Y.~H.}\ \bibnamefont {Chu}}, \bibinfo {author} {\bibfnamefont {Q.}~\bibnamefont {He}}, \bibinfo {author} {\bibfnamefont {C.~H.}\ \bibnamefont {Yang}}, \bibinfo {author} {\bibfnamefont {P.}~\bibnamefont {Yu}}, \bibinfo {author} {\bibfnamefont {L.~W.}\ \bibnamefont {Martin}}, \bibinfo {author} {\bibfnamefont {P.}~\bibnamefont {Shafer}},\ and\ \bibinfo {author} {\bibfnamefont {R.}~\bibnamefont {Ramesh}},\ }\href {https://doi.org/10.1021/nl900723j} {\bibfield  {journal} {\bibinfo  {journal} {Nano Letters}\ }\textbf {\bibinfo {volume} {9}},\ \bibinfo {pages} {1726} (\bibinfo {year} {2009})}\BibitemShut {NoStop}%
\bibitem [{\citenamefont {Tang}\ \emph {et~al.}(2015)\citenamefont {Tang}, \citenamefont {Zhu}, \citenamefont {Ma}, \citenamefont {Borisevich}, \citenamefont {Morozovska}, \citenamefont {Eliseev}, \citenamefont {Wang}, \citenamefont {Wang}, \citenamefont {Xu}, \citenamefont {Zhang},\ and\ \citenamefont {Pennycook}}]{Tang2015}%
  \BibitemOpen
  \bibfield  {author} {\bibinfo {author} {\bibfnamefont {Y.~L.}\ \bibnamefont {Tang}}, \bibinfo {author} {\bibfnamefont {Y.~L.}\ \bibnamefont {Zhu}}, \bibinfo {author} {\bibfnamefont {X.~L.}\ \bibnamefont {Ma}}, \bibinfo {author} {\bibfnamefont {A.~Y.}\ \bibnamefont {Borisevich}}, \bibinfo {author} {\bibfnamefont {A.~N.}\ \bibnamefont {Morozovska}}, \bibinfo {author} {\bibfnamefont {E.~A.}\ \bibnamefont {Eliseev}}, \bibinfo {author} {\bibfnamefont {W.~Y.}\ \bibnamefont {Wang}}, \bibinfo {author} {\bibfnamefont {Y.~J.}\ \bibnamefont {Wang}}, \bibinfo {author} {\bibfnamefont {Y.~B.}\ \bibnamefont {Xu}}, \bibinfo {author} {\bibfnamefont {Z.~D.}\ \bibnamefont {Zhang}},\ and\ \bibinfo {author} {\bibfnamefont {S.~J.}\ \bibnamefont {Pennycook}},\ }\href {https://doi.org/10.1126/science.1259869} {\bibfield  {journal} {\bibinfo  {journal} {Science}\ }\textbf {\bibinfo {volume} {348}},\ \bibinfo {pages} {547} (\bibinfo {year} {2015})}\BibitemShut {NoStop}%
\bibitem [{\citenamefont {Yadav}\ \emph {et~al.}(2016)\citenamefont {Yadav}, \citenamefont {Nelson}, \citenamefont {Hsu}, \citenamefont {Hong}, \citenamefont {Clarkson}, \citenamefont {Schlepuëtz}, \citenamefont {Damodaran}, \citenamefont {Shafer}, \citenamefont {Arenholz}, \citenamefont {Dedon}, \citenamefont {Chen}, \citenamefont {Vishwanath}, \citenamefont {Minor}, \citenamefont {Chen}, \citenamefont {Scott}, \citenamefont {Martin},\ and\ \citenamefont {Ramesh}}]{Yadav2016}%
  \BibitemOpen
  \bibfield  {author} {\bibinfo {author} {\bibfnamefont {A.~K.}\ \bibnamefont {Yadav}}, \bibinfo {author} {\bibfnamefont {C.~T.}\ \bibnamefont {Nelson}}, \bibinfo {author} {\bibfnamefont {S.~L.}\ \bibnamefont {Hsu}}, \bibinfo {author} {\bibfnamefont {Z.}~\bibnamefont {Hong}}, \bibinfo {author} {\bibfnamefont {J.~D.}\ \bibnamefont {Clarkson}}, \bibinfo {author} {\bibfnamefont {C.~M.}\ \bibnamefont {Schlepuëtz}}, \bibinfo {author} {\bibfnamefont {A.~R.}\ \bibnamefont {Damodaran}}, \bibinfo {author} {\bibfnamefont {P.}~\bibnamefont {Shafer}}, \bibinfo {author} {\bibfnamefont {E.}~\bibnamefont {Arenholz}}, \bibinfo {author} {\bibfnamefont {L.~R.}\ \bibnamefont {Dedon}}, \bibinfo {author} {\bibfnamefont {D.}~\bibnamefont {Chen}}, \bibinfo {author} {\bibfnamefont {A.}~\bibnamefont {Vishwanath}}, \bibinfo {author} {\bibfnamefont {A.~M.}\ \bibnamefont {Minor}}, \bibinfo {author} {\bibfnamefont {L.~Q.}\ \bibnamefont {Chen}}, \bibinfo {author} {\bibfnamefont {J.~F.}\ \bibnamefont {Scott}}, \bibinfo {author}
  {\bibfnamefont {L.~W.}\ \bibnamefont {Martin}},\ and\ \bibinfo {author} {\bibfnamefont {R.}~\bibnamefont {Ramesh}},\ }\href {https://doi.org/10.1038/nature16463} {\bibfield  {journal} {\bibinfo  {journal} {Nature}\ }\textbf {\bibinfo {volume} {530}},\ \bibinfo {pages} {198} (\bibinfo {year} {2016})}\BibitemShut {NoStop}%
\bibitem [{\citenamefont {Hong}\ \emph {et~al.}(2017)\citenamefont {Hong}, \citenamefont {Damodaran}, \citenamefont {Xue}, \citenamefont {Hsu}, \citenamefont {Britson}, \citenamefont {Yadav}, \citenamefont {Nelson}, \citenamefont {Wang}, \citenamefont {Scott}, \citenamefont {Martin}, \citenamefont {Ramesh},\ and\ \citenamefont {Chen}}]{Hong2017}%
  \BibitemOpen
  \bibfield  {author} {\bibinfo {author} {\bibfnamefont {Z.}~\bibnamefont {Hong}}, \bibinfo {author} {\bibfnamefont {A.~R.}\ \bibnamefont {Damodaran}}, \bibinfo {author} {\bibfnamefont {F.}~\bibnamefont {Xue}}, \bibinfo {author} {\bibfnamefont {S.~L.}\ \bibnamefont {Hsu}}, \bibinfo {author} {\bibfnamefont {J.}~\bibnamefont {Britson}}, \bibinfo {author} {\bibfnamefont {A.~K.}\ \bibnamefont {Yadav}}, \bibinfo {author} {\bibfnamefont {C.~T.}\ \bibnamefont {Nelson}}, \bibinfo {author} {\bibfnamefont {J.~J.}\ \bibnamefont {Wang}}, \bibinfo {author} {\bibfnamefont {J.~F.}\ \bibnamefont {Scott}}, \bibinfo {author} {\bibfnamefont {L.~W.}\ \bibnamefont {Martin}}, \bibinfo {author} {\bibfnamefont {R.}~\bibnamefont {Ramesh}},\ and\ \bibinfo {author} {\bibfnamefont {L.~Q.}\ \bibnamefont {Chen}},\ }\href {https://doi.org/10.1021/acs.nanolett.6b04875} {\bibfield  {journal} {\bibinfo  {journal} {Nano Letters}\ }\textbf {\bibinfo {volume} {17}},\ \bibinfo {pages} {2246} (\bibinfo {year} {2017})}\BibitemShut {NoStop}%
\bibitem [{\citenamefont {Das}\ \emph {et~al.}(2019)\citenamefont {Das}, \citenamefont {Tang}, \citenamefont {Hong}, \citenamefont {Gonçalves}, \citenamefont {McCarter}, \citenamefont {Klewe}, \citenamefont {Nguyen}, \citenamefont {{Gómez-Ortiz}}, \citenamefont {Shafer}, \citenamefont {Arenholz}, \citenamefont {Stoica}, \citenamefont {Hsu}, \citenamefont {Wang}, \citenamefont {Ophus}, \citenamefont {Liu}, \citenamefont {Nelson}, \citenamefont {Saremi}, \citenamefont {Prasad}, \citenamefont {Mei}, \citenamefont {Schlom}, \citenamefont {Íñiguez}, \citenamefont {{García-Fernández}}, \citenamefont {Muller}, \citenamefont {Chen}, \citenamefont {Junquera}, \citenamefont {Martin},\ and\ \citenamefont {Ramesh}}]{Das2019a}%
  \BibitemOpen
  \bibfield  {author} {\bibinfo {author} {\bibfnamefont {S.}~\bibnamefont {Das}}, \bibinfo {author} {\bibfnamefont {Y.~L.}\ \bibnamefont {Tang}}, \bibinfo {author} {\bibfnamefont {Z.}~\bibnamefont {Hong}}, \bibinfo {author} {\bibfnamefont {M.~A.}\ \bibnamefont {Gonçalves}}, \bibinfo {author} {\bibfnamefont {M.~R.}\ \bibnamefont {McCarter}}, \bibinfo {author} {\bibfnamefont {C.}~\bibnamefont {Klewe}}, \bibinfo {author} {\bibfnamefont {K.~X.}\ \bibnamefont {Nguyen}}, \bibinfo {author} {\bibfnamefont {F.}~\bibnamefont {{Gómez-Ortiz}}}, \bibinfo {author} {\bibfnamefont {P.}~\bibnamefont {Shafer}}, \bibinfo {author} {\bibfnamefont {E.}~\bibnamefont {Arenholz}}, \bibinfo {author} {\bibfnamefont {V.~A.}\ \bibnamefont {Stoica}}, \bibinfo {author} {\bibfnamefont {S.~L.}\ \bibnamefont {Hsu}}, \bibinfo {author} {\bibfnamefont {B.}~\bibnamefont {Wang}}, \bibinfo {author} {\bibfnamefont {C.}~\bibnamefont {Ophus}}, \bibinfo {author} {\bibfnamefont {J.~F.}\ \bibnamefont {Liu}}, \bibinfo {author} {\bibfnamefont {C.~T.}\
  \bibnamefont {Nelson}}, \bibinfo {author} {\bibfnamefont {S.}~\bibnamefont {Saremi}}, \bibinfo {author} {\bibfnamefont {B.}~\bibnamefont {Prasad}}, \bibinfo {author} {\bibfnamefont {A.~B.}\ \bibnamefont {Mei}}, \bibinfo {author} {\bibfnamefont {D.~G.}\ \bibnamefont {Schlom}}, \bibinfo {author} {\bibfnamefont {J.}~\bibnamefont {Íñiguez}}, \bibinfo {author} {\bibfnamefont {P.}~\bibnamefont {{García-Fernández}}}, \bibinfo {author} {\bibfnamefont {D.~A.}\ \bibnamefont {Muller}}, \bibinfo {author} {\bibfnamefont {L.~Q.}\ \bibnamefont {Chen}}, \bibinfo {author} {\bibfnamefont {J.}~\bibnamefont {Junquera}}, \bibinfo {author} {\bibfnamefont {L.~W.}\ \bibnamefont {Martin}},\ and\ \bibinfo {author} {\bibfnamefont {R.}~\bibnamefont {Ramesh}},\ }\href {https://doi.org/10.1038/s41586-019-1092-8} {\bibfield  {journal} {\bibinfo  {journal} {Nature}\ }\textbf {\bibinfo {volume} {568}},\ \bibinfo {pages} {368} (\bibinfo {year} {2019})}\BibitemShut {NoStop}%
\bibitem [{\citenamefont {Yin}\ \emph {et~al.}(2024)\citenamefont {Yin}, \citenamefont {Li}, \citenamefont {Zatterin}, \citenamefont {Rusu}, \citenamefont {Stylianidis}, \citenamefont {Hadjimichael}, \citenamefont {Aramberri}, \citenamefont {Iñiguez‐González}, \citenamefont {Conroy},\ and\ \citenamefont {Zubko}}]{yin2024}%
  \BibitemOpen
  \bibfield  {author} {\bibinfo {author} {\bibfnamefont {C.}~\bibnamefont {Yin}}, \bibinfo {author} {\bibfnamefont {Y.}~\bibnamefont {Li}}, \bibinfo {author} {\bibfnamefont {E.}~\bibnamefont {Zatterin}}, \bibinfo {author} {\bibfnamefont {D.}~\bibnamefont {Rusu}}, \bibinfo {author} {\bibfnamefont {E.}~\bibnamefont {Stylianidis}}, \bibinfo {author} {\bibfnamefont {M.}~\bibnamefont {Hadjimichael}}, \bibinfo {author} {\bibfnamefont {H.}~\bibnamefont {Aramberri}}, \bibinfo {author} {\bibfnamefont {J.}~\bibnamefont {Iñiguez‐González}}, \bibinfo {author} {\bibfnamefont {M.}~\bibnamefont {Conroy}},\ and\ \bibinfo {author} {\bibfnamefont {P.}~\bibnamefont {Zubko}},\ }\href {https://doi.org/10.1002/adma.202403985} {\bibfield  {journal} {\bibinfo  {journal} {Advanced Materials}\ }\textbf {\bibinfo {volume} {36}},\ \bibinfo {pages} {2403985} (\bibinfo {year} {2024})}\BibitemShut {NoStop}%
\bibitem [{\citenamefont {Rijnders}\ \emph {et~al.}(2004)\citenamefont {Rijnders}, \citenamefont {Blank}, \citenamefont {Choi},\ and\ \citenamefont {Eom}}]{rijnders04}%
  \BibitemOpen
  \bibfield  {author} {\bibinfo {author} {\bibfnamefont {G.}~\bibnamefont {Rijnders}}, \bibinfo {author} {\bibfnamefont {D.~H.}\ \bibnamefont {Blank}}, \bibinfo {author} {\bibfnamefont {J.}~\bibnamefont {Choi}},\ and\ \bibinfo {author} {\bibfnamefont {C.~B.}\ \bibnamefont {Eom}},\ }\href {https://doi.org/10.1063/1.1640472} {\bibfield  {journal} {\bibinfo  {journal} {Applied Physics Letters}\ }\textbf {\bibinfo {volume} {84}},\ \bibinfo {pages} {505} (\bibinfo {year} {2004})}\BibitemShut {NoStop}%
\bibitem [{\citenamefont {De~Luca}\ \emph {et~al.}(2017)\citenamefont {De~Luca}, \citenamefont {Strkalj}, \citenamefont {Manz}, \citenamefont {Bouillet}, \citenamefont {Fiebig},\ and\ \citenamefont {Trassin}}]{deluca2017a}%
  \BibitemOpen
  \bibfield  {author} {\bibinfo {author} {\bibfnamefont {G.}~\bibnamefont {De~Luca}}, \bibinfo {author} {\bibfnamefont {N.}~\bibnamefont {Strkalj}}, \bibinfo {author} {\bibfnamefont {S.}~\bibnamefont {Manz}}, \bibinfo {author} {\bibfnamefont {C.}~\bibnamefont {Bouillet}}, \bibinfo {author} {\bibfnamefont {M.}~\bibnamefont {Fiebig}},\ and\ \bibinfo {author} {\bibfnamefont {M.}~\bibnamefont {Trassin}},\ }\href {https://doi.org/10.1038/s41467-017-01620-2} {\bibfield  {journal} {\bibinfo  {journal} {Nature Communications}\ }\textbf {\bibinfo {volume} {8}},\ \bibinfo {pages} {1419} (\bibinfo {year} {2017})}\BibitemShut {NoStop}%
\bibitem [{\citenamefont {Gradauskaite}\ \emph {et~al.}(2024)\citenamefont {Gradauskaite}, \citenamefont {Yang}, \citenamefont {Efe}, \citenamefont {Pal}, \citenamefont {Fiebig},\ and\ \citenamefont {Trassin}}]{gradauskaite2024}%
  \BibitemOpen
  \bibfield  {author} {\bibinfo {author} {\bibfnamefont {E.}~\bibnamefont {Gradauskaite}}, \bibinfo {author} {\bibfnamefont {C.}~\bibnamefont {Yang}}, \bibinfo {author} {\bibfnamefont {I.}~\bibnamefont {Efe}}, \bibinfo {author} {\bibfnamefont {S.}~\bibnamefont {Pal}}, \bibinfo {author} {\bibfnamefont {M.}~\bibnamefont {Fiebig}},\ and\ \bibinfo {author} {\bibfnamefont {M.}~\bibnamefont {Trassin}},\ }\href {https://doi.org/10.1002/adfm.202412831} {\bibfield  {journal} {\bibinfo  {journal} {Advanced Functional Materials}\ ,\ \bibinfo {pages} {2412831}} (\bibinfo {year} {2024})}\BibitemShut {NoStop}%
\bibitem [{\citenamefont {Gradauskaite}\ \emph {et~al.}(2023)\citenamefont {Gradauskaite}, \citenamefont {Meier}, \citenamefont {Gray}, \citenamefont {Sarott}, \citenamefont {Scharsach}, \citenamefont {Campanini}, \citenamefont {Moran}, \citenamefont {Vogel}, \citenamefont {{Del Cid-Ledezma}}, \citenamefont {Huey}, \citenamefont {Rossell}, \citenamefont {Fiebig},\ and\ \citenamefont {Trassin}}]{gradauskaite2023}%
  \BibitemOpen
  \bibfield  {author} {\bibinfo {author} {\bibfnamefont {E.}~\bibnamefont {Gradauskaite}}, \bibinfo {author} {\bibfnamefont {Q.~N.}\ \bibnamefont {Meier}}, \bibinfo {author} {\bibfnamefont {N.}~\bibnamefont {Gray}}, \bibinfo {author} {\bibfnamefont {M.~F.}\ \bibnamefont {Sarott}}, \bibinfo {author} {\bibfnamefont {T.}~\bibnamefont {Scharsach}}, \bibinfo {author} {\bibfnamefont {M.}~\bibnamefont {Campanini}}, \bibinfo {author} {\bibfnamefont {T.}~\bibnamefont {Moran}}, \bibinfo {author} {\bibfnamefont {A.}~\bibnamefont {Vogel}}, \bibinfo {author} {\bibfnamefont {K.}~\bibnamefont {{Del Cid-Ledezma}}}, \bibinfo {author} {\bibfnamefont {B.~D.}\ \bibnamefont {Huey}}, \bibinfo {author} {\bibfnamefont {M.~D.}\ \bibnamefont {Rossell}}, \bibinfo {author} {\bibfnamefont {M.}~\bibnamefont {Fiebig}},\ and\ \bibinfo {author} {\bibfnamefont {M.}~\bibnamefont {Trassin}},\ }\href {https://doi.org/10.1038/s41563-023-01674-2} {\bibfield  {journal} {\bibinfo  {journal} {Nature Materials}\ }\textbf {\bibinfo {volume} {22}},\
  \bibinfo {pages} {1492} (\bibinfo {year} {2023})}\BibitemShut {NoStop}%
\bibitem [{\citenamefont {Abdelsamie}\ \emph {et~al.}(2024)\citenamefont {Abdelsamie}, \citenamefont {Chaudron}, \citenamefont {Bouzehouane}, \citenamefont {Dufour}, \citenamefont {Finco}, \citenamefont {Carrétéro}, \citenamefont {Jacques}, \citenamefont {Fusil},\ and\ \citenamefont {Garcia}}]{abdelsamie2024}%
  \BibitemOpen
  \bibfield  {author} {\bibinfo {author} {\bibfnamefont {A.}~\bibnamefont {Abdelsamie}}, \bibinfo {author} {\bibfnamefont {A.}~\bibnamefont {Chaudron}}, \bibinfo {author} {\bibfnamefont {K.}~\bibnamefont {Bouzehouane}}, \bibinfo {author} {\bibfnamefont {P.}~\bibnamefont {Dufour}}, \bibinfo {author} {\bibfnamefont {A.}~\bibnamefont {Finco}}, \bibinfo {author} {\bibfnamefont {C.}~\bibnamefont {Carrétéro}}, \bibinfo {author} {\bibfnamefont {V.}~\bibnamefont {Jacques}}, \bibinfo {author} {\bibfnamefont {S.}~\bibnamefont {Fusil}},\ and\ \bibinfo {author} {\bibfnamefont {V.}~\bibnamefont {Garcia}},\ }\href {https://doi.org/10.1063/5.0208996} {\bibfield  {journal} {\bibinfo  {journal} {Applied Physics Letters}\ }\textbf {\bibinfo {volume} {124}},\ \bibinfo {pages} {242902} (\bibinfo {year} {2024})}\BibitemShut {NoStop}%
\bibitem [{\citenamefont {Campanini}\ \emph {et~al.}(2020)\citenamefont {Campanini}, \citenamefont {Gradauskaite}, \citenamefont {Trassin}, \citenamefont {Yi}, \citenamefont {Yu}, \citenamefont {Ramesh}, \citenamefont {Erni},\ and\ \citenamefont {Rossell}}]{campanini2020a}%
  \BibitemOpen
  \bibfield  {author} {\bibinfo {author} {\bibfnamefont {M.}~\bibnamefont {Campanini}}, \bibinfo {author} {\bibfnamefont {E.}~\bibnamefont {Gradauskaite}}, \bibinfo {author} {\bibfnamefont {M.}~\bibnamefont {Trassin}}, \bibinfo {author} {\bibfnamefont {D.}~\bibnamefont {Yi}}, \bibinfo {author} {\bibfnamefont {P.}~\bibnamefont {Yu}}, \bibinfo {author} {\bibfnamefont {R.}~\bibnamefont {Ramesh}}, \bibinfo {author} {\bibfnamefont {R.}~\bibnamefont {Erni}},\ and\ \bibinfo {author} {\bibfnamefont {M.~D.}\ \bibnamefont {Rossell}},\ }\href {https://doi.org/10.1039/D0NR01258K} {\bibfield  {journal} {\bibinfo  {journal} {Nanoscale}\ }\textbf {\bibinfo {volume} {12}},\ \bibinfo {pages} {9186} (\bibinfo {year} {2020})}\BibitemShut {NoStop}%
\bibitem [{\citenamefont {Nataf}\ \emph {et~al.}(2020)\citenamefont {Nataf}, \citenamefont {Guennou}, \citenamefont {Gregg}, \citenamefont {Meier}, \citenamefont {Hlinka}, \citenamefont {Salje},\ and\ \citenamefont {Kreisel}}]{Nataf2020}%
  \BibitemOpen
  \bibfield  {author} {\bibinfo {author} {\bibfnamefont {G.~F.}\ \bibnamefont {Nataf}}, \bibinfo {author} {\bibfnamefont {M.}~\bibnamefont {Guennou}}, \bibinfo {author} {\bibfnamefont {J.~M.}\ \bibnamefont {Gregg}}, \bibinfo {author} {\bibfnamefont {D.}~\bibnamefont {Meier}}, \bibinfo {author} {\bibfnamefont {J.}~\bibnamefont {Hlinka}}, \bibinfo {author} {\bibfnamefont {E.~K.}\ \bibnamefont {Salje}},\ and\ \bibinfo {author} {\bibfnamefont {J.}~\bibnamefont {Kreisel}},\ }\href {https://doi.org/10.1038/s42254-020-0235-z} {\bibfield  {journal} {\bibinfo  {journal} {Nature Reviews Physics}\ }\textbf {\bibinfo {volume} {2}},\ \bibinfo {pages} {634–648} (\bibinfo {year} {2020})}\BibitemShut {NoStop}%
\bibitem [{\citenamefont {Jin}\ \emph {et~al.}(2017)\citenamefont {Jin}, \citenamefont {Xu}, \citenamefont {Zeng}, \citenamefont {Lu}, \citenamefont {Barthel}, \citenamefont {Schulthess}, \citenamefont {{Dunin-Borkowski}}, \citenamefont {Wang},\ and\ \citenamefont {Jia}}]{jin2017b}%
  \BibitemOpen
  \bibfield  {author} {\bibinfo {author} {\bibfnamefont {L.}~\bibnamefont {Jin}}, \bibinfo {author} {\bibfnamefont {P.~X.}\ \bibnamefont {Xu}}, \bibinfo {author} {\bibfnamefont {Y.}~\bibnamefont {Zeng}}, \bibinfo {author} {\bibfnamefont {L.}~\bibnamefont {Lu}}, \bibinfo {author} {\bibfnamefont {J.}~\bibnamefont {Barthel}}, \bibinfo {author} {\bibfnamefont {T.}~\bibnamefont {Schulthess}}, \bibinfo {author} {\bibfnamefont {R.~E.}\ \bibnamefont {{Dunin-Borkowski}}}, \bibinfo {author} {\bibfnamefont {H.}~\bibnamefont {Wang}},\ and\ \bibinfo {author} {\bibfnamefont {C.~L.}\ \bibnamefont {Jia}},\ }\href {https://doi.org/10.1038/srep39698} {\bibfield  {journal} {\bibinfo  {journal} {Scientific Reports}\ }\textbf {\bibinfo {volume} {7}},\ \bibinfo {pages} {39698} (\bibinfo {year} {2017})}\BibitemShut {NoStop}%
\bibitem [{\citenamefont {Spaldin}\ \emph {et~al.}(2021)\citenamefont {Spaldin}, \citenamefont {Efe}, \citenamefont {Rossell},\ and\ \citenamefont {Gattinoni}}]{Spaldin2021}%
  \BibitemOpen
  \bibfield  {author} {\bibinfo {author} {\bibfnamefont {N.~A.}\ \bibnamefont {Spaldin}}, \bibinfo {author} {\bibfnamefont {I.}~\bibnamefont {Efe}}, \bibinfo {author} {\bibfnamefont {M.~D.}\ \bibnamefont {Rossell}},\ and\ \bibinfo {author} {\bibfnamefont {C.}~\bibnamefont {Gattinoni}},\ }\href {https://doi.org/10.1063/5.0046061} {\bibfield  {journal} {\bibinfo  {journal} {The Journal of Chemical Physics}\ }\textbf {\bibinfo {volume} {154}},\ \bibinfo {pages} {154702} (\bibinfo {year} {2021})}\BibitemShut {NoStop}%
\bibitem [{\citenamefont {Xie}\ \emph {et~al.}(2017)\citenamefont {Xie}, \citenamefont {Li}, \citenamefont {Heikes}, \citenamefont {Zhang}, \citenamefont {Hong}, \citenamefont {Gao}, \citenamefont {Nelson}, \citenamefont {Xue}, \citenamefont {Kioupakis}, \citenamefont {Chen}, \citenamefont {Schlom}, \citenamefont {Wang},\ and\ \citenamefont {Pan}}]{Xie2017}%
  \BibitemOpen
  \bibfield  {author} {\bibinfo {author} {\bibfnamefont {L.}~\bibnamefont {Xie}}, \bibinfo {author} {\bibfnamefont {L.}~\bibnamefont {Li}}, \bibinfo {author} {\bibfnamefont {C.~A.}\ \bibnamefont {Heikes}}, \bibinfo {author} {\bibfnamefont {Y.}~\bibnamefont {Zhang}}, \bibinfo {author} {\bibfnamefont {Z.}~\bibnamefont {Hong}}, \bibinfo {author} {\bibfnamefont {P.}~\bibnamefont {Gao}}, \bibinfo {author} {\bibfnamefont {C.~T.}\ \bibnamefont {Nelson}}, \bibinfo {author} {\bibfnamefont {F.}~\bibnamefont {Xue}}, \bibinfo {author} {\bibfnamefont {E.}~\bibnamefont {Kioupakis}}, \bibinfo {author} {\bibfnamefont {L.}~\bibnamefont {Chen}}, \bibinfo {author} {\bibfnamefont {D.~G.}\ \bibnamefont {Schlom}}, \bibinfo {author} {\bibfnamefont {P.}~\bibnamefont {Wang}},\ and\ \bibinfo {author} {\bibfnamefont {X.}~\bibnamefont {Pan}},\ }\href {https://doi.org/10.1002/adma.201701475} {\bibfield  {journal} {\bibinfo  {journal} {Advanced Materials}\ }\textbf {\bibinfo {volume} {29}},\ \bibinfo {pages} {1701475} (\bibinfo {year}
  {2017})}\BibitemShut {NoStop}%
\bibitem [{\citenamefont {Efe}\ \emph {et~al.}(2025)\citenamefont {Efe}, \citenamefont {Vogel}, \citenamefont {Huxter}, \citenamefont {Gradauskaite}, \citenamefont {Gaponenko}, \citenamefont {Paruch}, \citenamefont {Degen}, \citenamefont {Rossell}, \citenamefont {Fiebig},\ and\ \citenamefont {Trassin}}]{efe2025}%
  \BibitemOpen
  \bibfield  {author} {\bibinfo {author} {\bibfnamefont {I.}~\bibnamefont {Efe}}, \bibinfo {author} {\bibfnamefont {A.}~\bibnamefont {Vogel}}, \bibinfo {author} {\bibfnamefont {W.~S.}\ \bibnamefont {Huxter}}, \bibinfo {author} {\bibfnamefont {E.}~\bibnamefont {Gradauskaite}}, \bibinfo {author} {\bibfnamefont {I.}~\bibnamefont {Gaponenko}}, \bibinfo {author} {\bibfnamefont {P.}~\bibnamefont {Paruch}}, \bibinfo {author} {\bibfnamefont {C.~L.}\ \bibnamefont {Degen}}, \bibinfo {author} {\bibfnamefont {M.~D.}\ \bibnamefont {Rossell}}, \bibinfo {author} {\bibfnamefont {M.}~\bibnamefont {Fiebig}},\ and\ \bibinfo {author} {\bibfnamefont {M.}~\bibnamefont {Trassin}},\ }\href {https://doi.org/10.1038/s41467-025-60176-8} {\bibfield  {journal} {\bibinfo  {journal} {Nature Communications}\ }\textbf {\bibinfo {volume} {16}},\ \bibinfo {pages} {6131} (\bibinfo {year} {2025})}\BibitemShut {NoStop}%
\bibitem [{\citenamefont {Aurivillius}(1949)}]{aurivilliusmain}%
  \BibitemOpen
  \bibfield  {author} {\bibinfo {author} {\bibfnamefont {B.}~\bibnamefont {Aurivillius}},\ }\href@noop {} {\bibfield  {journal} {\bibinfo  {journal} {Arkiv. Kemi.}\ }\textbf {\bibinfo {volume} {1}},\ \bibinfo {pages} {463} (\bibinfo {year} {1949})}\BibitemShut {NoStop}%
\bibitem [{\citenamefont {Moure}(2018)}]{Moure2018a}%
  \BibitemOpen
  \bibfield  {author} {\bibinfo {author} {\bibfnamefont {A.}~\bibnamefont {Moure}},\ }\href {https://doi.org/10.3390/app8010062} {\bibfield  {journal} {\bibinfo  {journal} {Applied Sciences}\ }\textbf {\bibinfo {volume} {8}},\ \bibinfo {pages} {62} (\bibinfo {year} {2018})}\BibitemShut {NoStop}%
\bibitem [{\citenamefont {Gradauskaite}\ \emph {et~al.}(2020)\citenamefont {Gradauskaite}, \citenamefont {Campanini}, \citenamefont {Biswas}, \citenamefont {Schneider}, \citenamefont {Fiebig}, \citenamefont {Rossell},\ and\ \citenamefont {Trassin}}]{gradauskaite2020}%
  \BibitemOpen
  \bibfield  {author} {\bibinfo {author} {\bibfnamefont {E.}~\bibnamefont {Gradauskaite}}, \bibinfo {author} {\bibfnamefont {M.}~\bibnamefont {Campanini}}, \bibinfo {author} {\bibfnamefont {B.}~\bibnamefont {Biswas}}, \bibinfo {author} {\bibfnamefont {C.~W.}\ \bibnamefont {Schneider}}, \bibinfo {author} {\bibfnamefont {M.}~\bibnamefont {Fiebig}}, \bibinfo {author} {\bibfnamefont {M.~D.}\ \bibnamefont {Rossell}},\ and\ \bibinfo {author} {\bibfnamefont {M.}~\bibnamefont {Trassin}},\ }\href {https://doi.org/10.1002/admi.202000202} {\bibfield  {journal} {\bibinfo  {journal} {Advanced Materials Interfaces}\ }\textbf {\bibinfo {volume} {7}},\ \bibinfo {pages} {2000202} (\bibinfo {year} {2020})}\BibitemShut {NoStop}%
\bibitem [{\citenamefont {Gradauskaite}\ \emph {et~al.}(2021)\citenamefont {Gradauskaite}, \citenamefont {Gray}, \citenamefont {Campanini}, \citenamefont {Rossell},\ and\ \citenamefont {Trassin}}]{Gradauskaite2021a}%
  \BibitemOpen
  \bibfield  {author} {\bibinfo {author} {\bibfnamefont {E.}~\bibnamefont {Gradauskaite}}, \bibinfo {author} {\bibfnamefont {N.}~\bibnamefont {Gray}}, \bibinfo {author} {\bibfnamefont {M.}~\bibnamefont {Campanini}}, \bibinfo {author} {\bibfnamefont {M.~D.}\ \bibnamefont {Rossell}},\ and\ \bibinfo {author} {\bibfnamefont {M.}~\bibnamefont {Trassin}},\ }\href {https://doi.org/10.1021/acs.chemmater.1c03466} {\bibfield  {journal} {\bibinfo  {journal} {Chemistry of Materials}\ }\textbf {\bibinfo {volume} {33}},\ \bibinfo {pages} {9439} (\bibinfo {year} {2021})}\BibitemShut {NoStop}%
\bibitem [{\citenamefont {Zhao}\ \emph {et~al.}(2021)\citenamefont {Zhao}, \citenamefont {Chen}, \citenamefont {Prosandeev}, \citenamefont {Artyukhin},\ and\ \citenamefont {Bellaiche}}]{Zhao2021}%
  \BibitemOpen
  \bibfield  {author} {\bibinfo {author} {\bibfnamefont {H.~J.}\ \bibnamefont {Zhao}}, \bibinfo {author} {\bibfnamefont {P.}~\bibnamefont {Chen}}, \bibinfo {author} {\bibfnamefont {S.}~\bibnamefont {Prosandeev}}, \bibinfo {author} {\bibfnamefont {S.}~\bibnamefont {Artyukhin}},\ and\ \bibinfo {author} {\bibfnamefont {L.}~\bibnamefont {Bellaiche}},\ }\href {https://doi.org/10.1038/s41563-020-00821-3} {\bibfield  {journal} {\bibinfo  {journal} {Nature Materials}\ }\textbf {\bibinfo {volume} {20}},\ \bibinfo {pages} {341} (\bibinfo {year} {2021})}\BibitemShut {NoStop}%
\bibitem [{\citenamefont {Campanini}\ \emph {et~al.}(2018)\citenamefont {Campanini}, \citenamefont {Erni}, \citenamefont {Yang}, \citenamefont {Ramesh},\ and\ \citenamefont {Rossell}}]{Campanini2018}%
  \BibitemOpen
  \bibfield  {author} {\bibinfo {author} {\bibfnamefont {M.}~\bibnamefont {Campanini}}, \bibinfo {author} {\bibfnamefont {R.}~\bibnamefont {Erni}}, \bibinfo {author} {\bibfnamefont {C.~H.}\ \bibnamefont {Yang}}, \bibinfo {author} {\bibfnamefont {R.}~\bibnamefont {Ramesh}},\ and\ \bibinfo {author} {\bibfnamefont {M.~D.}\ \bibnamefont {Rossell}},\ }\href {https://doi.org/10.1021/acs.nanolett.7b03817} {\bibfield  {journal} {\bibinfo  {journal} {Nano Letters}\ }\textbf {\bibinfo {volume} {18}},\ \bibinfo {pages} {717} (\bibinfo {year} {2018})}\BibitemShut {NoStop}%
\end{thebibliography}%

\appendix
\newpage
\section*{Supporting Information}

\renewcommand{\thefigure}{S\arabic{figure}}
\renewcommand{\thetable}{S\arabic{table}}
\setcounter{figure}{0} 
\setcounter{table}{0}  

\subsection*{Supporting Note S1: Uniaxial in-plane polarization components in BFO films on LSMO-buffered NGO (001) confirmed with vector PFM measurements }

\begin{figure}[htb!]
  \centering 
  \begin{adjustbox}{width=0.9\textwidth, center}
    \includegraphics[width=0.9\textwidth, clip, trim=8 8 8 8]{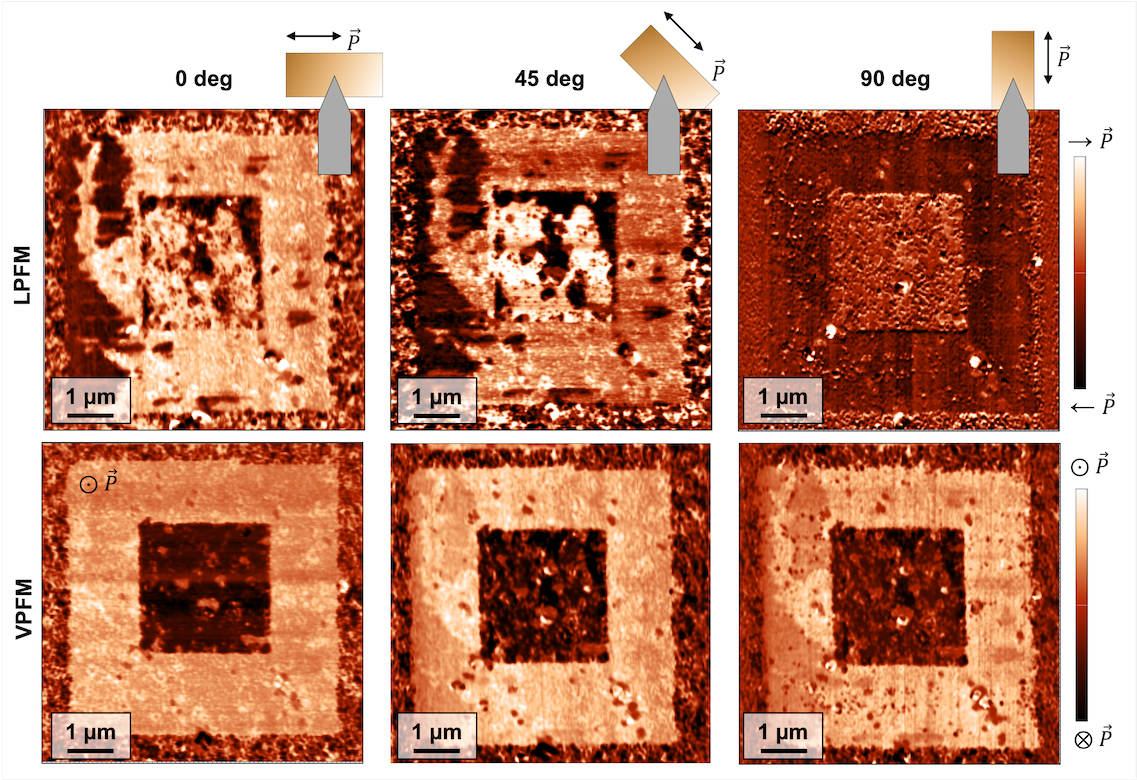}
  \end{adjustbox}
   \caption{\textbf{Vector PFM measurements confirming uniaxial in-plane polarization components in 25-nm-thick BFO films grown on LSMO-buffered NGO (001).}  
  (a) At 0° sample rotation, the in-plane polarization (\(\parallel\) BFO [110]\textsubscript{pc}) is perpendicular to the cantilever, producing a maximum torsional response recorded in the LPFM channel.  
  (b) At 45° rotation, the polarization projection has equal components transverse and longitudinal to the cantilever, yielding mixed torsion (LPFM) and buckling (VPFM) signals and a reduced LPFM amplitude.  
  (c) At 90° rotation (cantilever \(\parallel\) BFO [010]\textsubscript{o}), the in-plane polarization aligns parallel to the cantilever, shifting the response entirely into the VPFM (buckling) channel.  
  This series of sample-rotation-dependent PFM measurements demonstrates that the BFO in-plane polarization components are uniaxial and oriented along the NGO [010]\textsubscript{o} direction.}
\end{figure}

 \newpage

 \subsection*{Supporting Note S2: Uniaxial in-plane polarization components in BFO films on BFTO (1 u.c.)/LSMO-buffered NGO (001) confirmed with vector PFM measurements}

\begin{figure}[htb!]
  \centering 
  \begin{adjustbox}{width=0.9\textwidth, center}
    \includegraphics[width=0.9\textwidth, clip, trim=8 8 8 8]{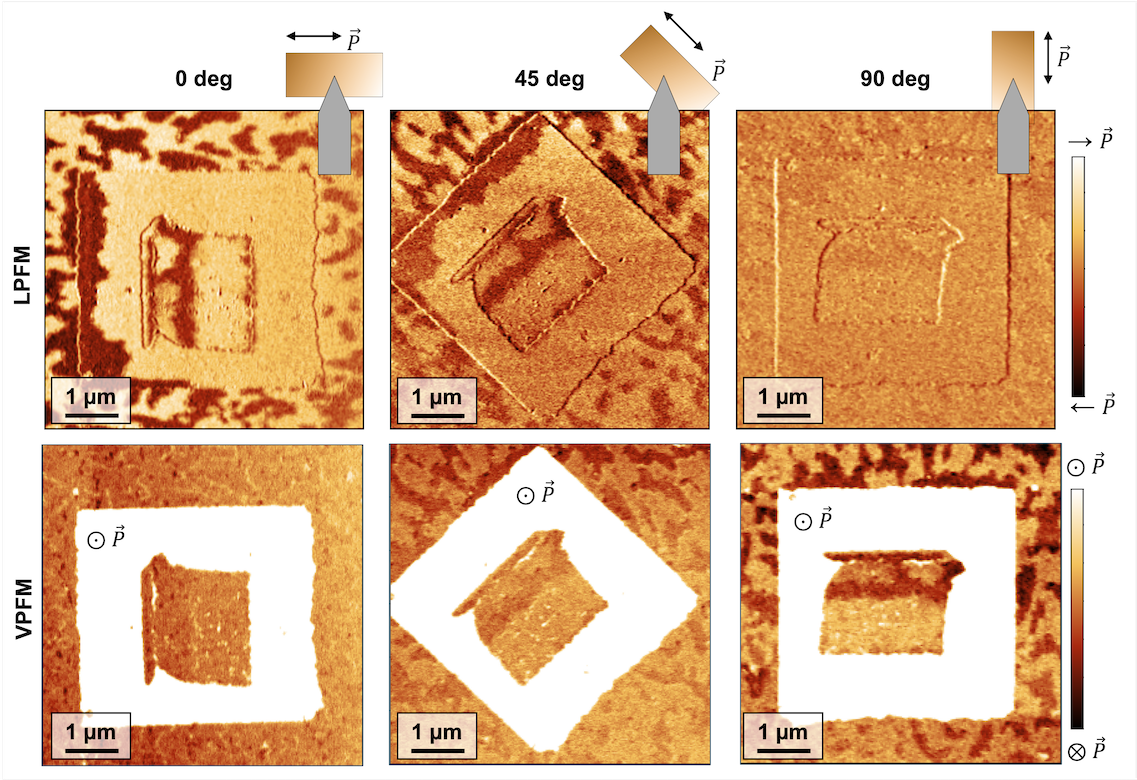}
  \end{adjustbox}
  \caption{\textbf{Vector PFM rotation series confirming uniaxial in-plane polarization in 25-nm-thick BFO films grown on BFTO (1 u.c.)/LSMO-buffered NGO (001).}  
  (a) At 0° sample rotation (cantilever \(\parallel\) BFO [110]\textsubscript{pc}), the in-plane polarization is perpendicular to the cantilever, yielding a maximum torsional response in the LPFM channel.  
  (b) At 45° rotation, equal polarization components transverse and longitudinal to the cantilever produce mixed torsion (LPFM) and buckling (VPFM) signals, reducing the LPFM amplitude.  
  (c) At 90° rotation (cantilever \(\parallel\) BFO [010]\textsubscript{o}), the in-plane polarization aligns parallel to the cantilever, shifting the response entirely into the VPFM (buckling) channel. The VPFM scale bar is adjusted here to highlight domains that appear in the downward-poled regions due to cantilever buckling.  
  This series of sample-rotation-dependent PFM measurements demonstrates that the BFO in-plane polarization components are uniaxial and remain oriented along the NGO [010]\textsubscript{o} direction after 1 u.c. of BFTO is inserted below the BFO film.}
  \label{fig:S2}
\end{figure}

\newpage

\subsection*{Supporting Note S3: Polar homochiral Néel domain walls in pristine BFO films on BFTO (1 u.c.)/LSMO/NGO(001)}

\begin{figure}[htb!]
  \centering
  \includegraphics[width=0.5\textwidth, clip, trim=8 8 8 8]{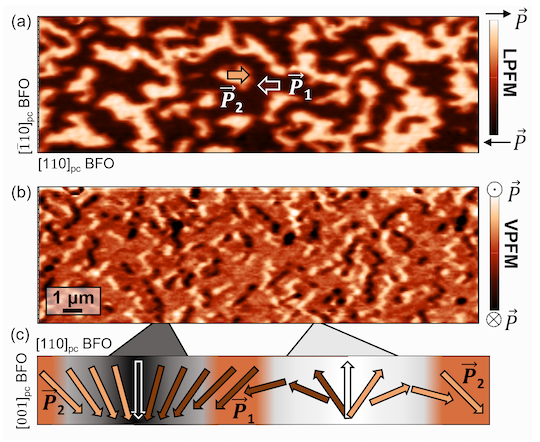}
 \caption{\textbf{Polar homochiral Néel domain walls in pristine 19-nm BFO films grown on BFTO (1 u.c.)/LSMO/NGO(001) consistent with a previous report\cite{gradauskaite2023}.}  
  \textbf{a}, LPFM phase image showing only two in-plane polarization variants, [$11\bar{1}$] ($\vec{P}_2$) and [$\bar{1}\bar{1}\bar{1}$] ($\vec{P}_1$), in the as-deposited film.  
  \textbf{b}, VPFM phase image of the same area, revealing a uniform downward out-of-plane polarization with local discontinuities at domain walls that appear as alternating dark (downward) and bright (upward) lines, indicating a consistent sense of polarization rotation or homochirality across all walls\cite{gradauskaite2023}. 
  \textbf{c}, Schematic of homochiral head-to-head and tail-to-tail Néel walls stabilized by the in-plane polarization of the BFTO buffer, illustrating the uniform rotational sense of the polarization vector across each wall in the cross-section of $(\overline{1}10)_{\mathrm{pc}}$ BFO plane.}
\end{figure}

\end{document}